\documentclass[12pt,preprint]{aastex}

\begin{document}

\shorttitle{Disk Evolution in Cepheus B}
\shortauthors{Getman et al.} 
\slugcomment{Accepted for publication in The Astrophysical Journal 04/30/09}

\title{Protoplanetary Disk Evolution around the Triggered Star Forming Region Cepheus B}

\author{Konstantin V.\ Getman\altaffilmark{1}, Eric D.\
Feigelson\altaffilmark{1,2}, Kevin L.\ Luhman\altaffilmark{1,2},
Aurora Sicilia-Aguilar\altaffilmark{3}, Junfeng Wang\altaffilmark{4}, Gordon P.\
Garmire\altaffilmark{1}}

\altaffiltext{1}{Department of Astronomy \& Astrophysics, 525
Davey Laboratory, Pennsylvania State University, University Park
PA 16802} \altaffiltext{2}{Center for Exoplanets and Habitable Worlds, 525
Davey Laboratory, Pennsylvania State University, University Park
PA 16802} \altaffiltext{3}{Max-Planck-Institut fur Astronomie, Konigstuhl 17, 69117 Heidelberg,
Germany} \altaffiltext{4}{Harvard-Smithsonian Center for Astrophysics, 60 Garden Street, Cambridge, MA 02138, USA}

\email{gkosta@astro.psu.edu}

\begin{abstract}
The Cepheus B (Cep~B) molecular cloud and a portion of the nearby Cep~OB3b OB association, one of the most active regions of star formation within 1~kpc, have been observed with the IRAC detector on board the {\it Spitzer Space Telescope}. The goals are to study protoplanetary disk evolution and processes of sequential triggered star formation in the region. Out of $\sim 400$ pre-main sequence (PMS) stars selected with an earlier {\it Chandra X-ray Observatory} observation, $\sim 95\%$ are identified with mid-infrared sources and most of these are classified as diskless or disk-bearing stars.  The discovery of the additional $> 200$ IR-excess low-mass members gives a combined {\it Chandra}$+${\it Spitzer} PMS sample that is almost complete down to $0.5$~M$_{\odot}$ outside of the cloud, and somewhat above $1$~M$_{\odot}$ in the cloud.

Analyses of the nearly disk-unbiased combined $Chandra$ and $Spitzer$ selected stellar sample give several results. Our major finding is a spatio-temporal gradient of young stars from the hot molecular core towards the primary ionizing O star HD 217086. This strongly supports the radiation driven implosion (RDI) model of triggered star formation in the region. The empirical estimate for the shock velocity of $\sim 1$~km/s is very similar to theoretical models of RDI in shocked molecular clouds. The initial mass function (IMF) of the lightly obscured triggered population exhibits a standard Galactic field IMF shape. The unusually high apparent value of $\ga 70\%$  star formation efficiency inferred from the ratio of star mass to current molecular gas mass indicates that most of the Cep~B molecular cloud has been already ablated or transformed to stars. Contrary to the current RDI simulations, our findings indicate that star formation triggering by HII region shocks is not restricted to a single episode but can continue for millions of years. Other results include: 1. agreement of the disk fractions, their mass dependency, and fractions of transition disks with other clusters; 2. confirmation of the youthfulness of the embedded Cep~B cluster; 3. confirmation of the effect of suppression of time-integrated X-ray emission in disk-bearing versus diskless systems. 
\end{abstract}


\section{INTRODUCTION \label{introduction_section}}

\begin{figure}
\centering
\includegraphics[angle=0.,width=7.0in]{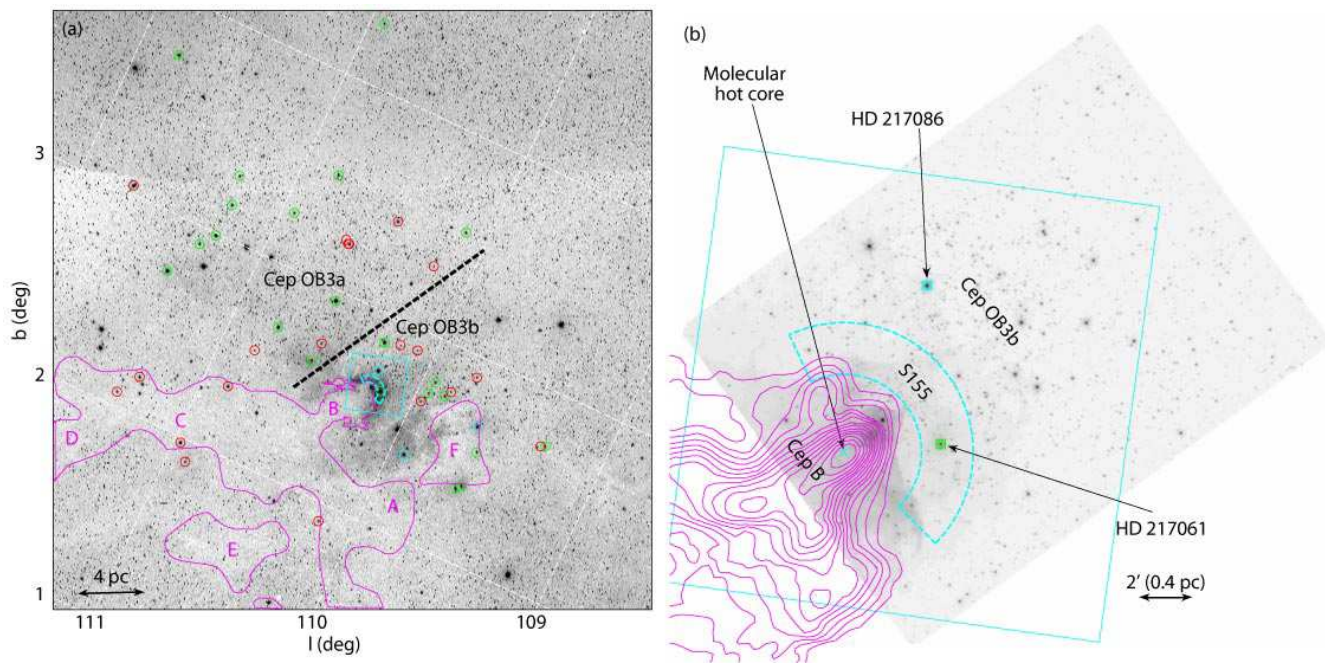}
\caption{(a) Large-scale optical image from the Digitized Sky Survey (DSS) of the Cepheus~B/OB3 environs. Magenta contours and core labels outline $^{12}$CO emission  from \citet{Sargent77}.  O stars (cyan circles), B0-B3 (green), B4-B9 (red) are from \citet{Blaauw59}. The black dashed line demarcates the older sub-group Cep~OB3a  from Cep~OB3b. North is up, east is to the left.  (b) Closeup of the Cep~B region showing the $20\arcmin \times 15\arcmin$ IRAC image in the 4.5~$\mu$m band. The ionizing sources of the region, the O7 star HD~217086 and B1 star HD~217061, are marked. The magenta contours show the $^{12}$CO emission measured by \citet{Beuther00} with the central hot core marked. The  $17\arcmin \times 17\arcmin$ $Chandra$ ACIS-I field is outlined in cyan. The H~II S~155 interface between the Cep~B molecular cloud and unobscured Cep~OB3b association is schematically indicated in cyan. \label{fig_map}}
\end{figure}

Cepheus~B (Cep~B) is a molecular core located at the edge of the Cepheus giant molecular cloud at a distance around 725~pc and lying $2.6^\circ$ above the Galactic Plane \citep{Sargent77,Yu96}. A handful of embedded young stars were found in Cep~B from radio continuum and infrared (IR) studies \citep{Felli78, Testi95}. The unobscured stellar OB association Cep~OB3  lies around Cep~B; the younger subgroup, Cep~OB3b, lies closest to the cloud as shown in Figure \ref{fig_map} \citep{Blaauw64,Kun08}. For many years, the Cep~OB3 association has been considered to be a good example of large-scale sequential star formation in accord with the model of \citet{Elmegreen77} where stellar winds and supernova remnants of an older stellar cluster compress and trigger a second generation of star formation in nearby molecular cloud cores \citep{Sargent79}.  

The interface between the molecular cloud and the Cep~OB3b star association is clearly delineated by the optically bright H{\sc II} region Sharpless 155 (S~155), where cloud material is ionized and heated by the radiation field of the O7 star HD~217086, B1 star HD~217061 and perhaps other cluster members \citep{Panagia81,Beuther00}. Figure \ref{fig_map} shows the spatial relationship of the cloud, H{\sc II} region, and exciting stars.  Unlike in the Orion Nebula where the H{\sc II} region lies between the cloud and ourselves, the photodissociation region at S~155 is favorably oriented to reveal the progression of star formation.  Following the triggered star formation model, we expect the surface of the cloud to be eroded by the early-type stars so that the cloud edge moves eastward across the observer's field of view with new stars emerging from the obscuring molecular cloud. Sources located within but near the edge of the molecular cloud would represent a new generation of star formation triggered by the H{\sc II} region shock propagating into the cloud.

This scenario of triggered star formation has been recently strengthened by the discovery using the {\it Chandra X-ray Observatory} of $> 300$ lightly-obscured low-mass pre-main sequence (PMS) members of the Cep~OB3b cluster located outside of the cloud, and a rich population of $\sim 60$ PMS stars in the cluster embedded within Cep~B \citep{Getman06}.  X-ray surveys are particularly effective in discriminating disk-free PMS populations from Galactic field stars which often badly contaminate infrared (IR) surveys of young stellar clusters \citep[see review by][]{Feigelson07}. Using 2MASS counterparts of the X-ray sources, \citet{Getman06} found that PMS stars in the embedded cluster are more likely (26\% $vs.$ 4\%) to have $K$-band excesses from heated inner protoplanetary disks than stars in the unobscured Cep~OB3 region. This supports both, youthfulness of the embedded Cep~B population and the prevalence of planet forming disks in the embedded cluster.

We seek here to elucidate the relationships of disks and environments in this region using mid-IR photometry from the {\it Spizer Space Telescope} which is exceptionally well-suited for detecting disks around low-mass members of young clusters \citep[e.g.][and references therein]{Allen04,Sicilia-Aguilar06,Luhman08b}.  Measures of disk fractions as functions of stellar mass and age in different star-forming environments can provide better understanding of the evolution of circumstellar disks in different environments and thus have direct astrophysical importance in evaluating the conditions for planet formation. X-ray PMS selection is very complementary to IR surveys. As X-ray emission from PMS stars is based on enhanced solar-type magnetic reconnection events rather than disk or accretion processes, X-ray selection delivers rich and clean samples of diskless stars missed by IR selection \citep{Feigelson07}. The X-ray stars identified by \citet{Getman06} in combination with discovered here $Spitzer$ disk-bearing stars constitute a valuable sample to measure protoplanetary disk properties in different radiative environments and to study the triggering process inside the Cep~B molecular cloud. We use the spatial distribution of disks around young stars of the selected $Chandra$ and $Spitzer$ sample to test concepts of cloud ablation and the radiative driven implosion (RDI) model of triggered star formation on the edges of H{\sc II} regions \citep[e.g.,][]{Bertoldi89}. 

In \S \ref{data_reduction_section} we report the $Spitzer$ observations using the InfraRed Array Camera (IRAC) detector. Classification on disk-bearing and diskless X-ray emitting PMS stars is provided in \S \ref{yso_classification_section} and the discovery of the new population of non-$Chandra$ IR-excess members is given in \S \ref{non_xray_subsection}. The IMF and mass completeness limits of the combined $Chandra$ and $Spizer$ stellar samples of different parts of the region are considered in \S \ref{IMF_subsection}. The results on the disk evolution of the selected PMS sample are presented in \S \ref{results_section}.  We end in \S \ref{discussion_section} with the implications of the new observational findings for the star formation process in the region.  We adopt a distance of 725~pc to the star forming complex and infer an age of $2-3$~Myr for the Cep~OB3b stars and an age between $<1$ and 2~Myr for the embedded Cep~B stars, as discussed in Appendix \ref{distance_age_subsection}. Appendix \ref{lx_disks_section} presents the relationship between X-ray activity, stellar mass and disks; these results confirm those found in other young stellar clusters.

\section{CHANDRA AND SPITZER OBSERVATIONS  \label{data_reduction_section}}

\subsection{X-ray Data \label{xray_data_subsection}}

\begin{figure}
\centering
\includegraphics[angle=0.,width=7.0in]{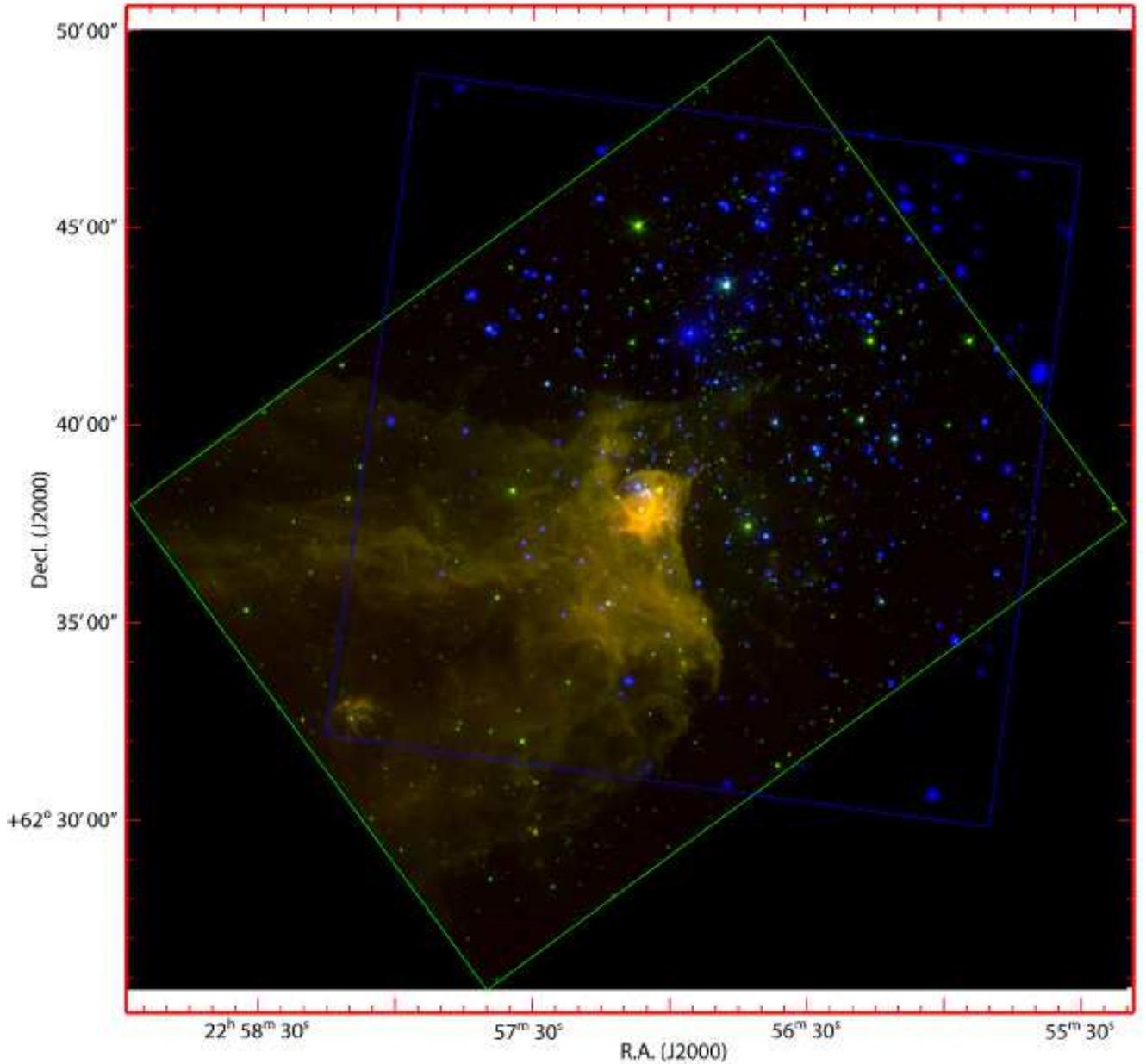}
\caption{Combined X-ray and IR image of the Cep~B/Cep~OB3b region. The adaptively smoothed $Chandra$ ACIS-I image in the $0.5-8.0$~keV band (blue) is superposed on the $Spitzer$ IRAC composite image in the 3.6~$\mu$m (green) and 5.8~$\mu$m (red) bands. The  $17\arcmin \times 17\arcmin$ $Chandra$ ACIS-I field is outlined in blue and the $\sim 20\arcmin \times 15\arcmin$ field of the $Spitzer$ IRAC mosaic in 3.6/5.8~$\mu$m bands is outlined in green. \label{fig_color_image}}
\end{figure}

The X-ray observations  of the Cep~B/OB3b and their data analysis are described in detail by \citet{Getman06}.  The 30~ks exposure was obtained on 2003 March $11.51-11.88$ with the Advanced CCD Imaging Spectrometer (ACIS) detector \citep{Garmire03} on-board NASA's {\it Chandra X-ray Observatory} \citep{Weisskopf02} as part of the ACIS Instrument Team's Guaranteed Time Observions (ObsId \#3502, G. Garmire, PI).  The imaging array (ACIS-I) consists of four abutted $1024 \times 1024$ pixel front-side illuminated charge-coupled devices (CCDs) covering about $17\arcmin \times 17 \arcmin$ on the sky.  Following data reduction based on the {\it ACIS Extract} IDL script, \citet{Getman06} obtained an X-ray catalog of 431 sources, most with $< 0.5\arcsec$ positional accuracy (Figure \ref{fig_color_image}). Using 2MASS counterparts, $89\%$ of the X-ray sources are confidently associated with cluster PMS members of the region. The remaining X-ray sources are mostly extragalactic active galactic nuclei.

\subsection{IRAC Data \label{irac_data_reduction_subsection}}

The mid-IR observation of Cep~B and Cep~OB3b was obtained on 2007 February 18  with the IRAC detector \citep{Fazio04} on NASA's {\it Spitzer Space Telescope} in the 3.6, 4.5, 5.8, and 8.0~$\mu$m channels. This is a General Observer project (program identification \#30361; J. Wang, PI). Two adjacent fields subtending $\sim 20\arcmin \times 15\arcmin$ in channel pairs 3.6/5.8~$\mu$m and 4.5/8.0~$\mu$m were centered, respectively, on $\alpha = 22^{\rm{h}}57^{\rm{m}}09\fs7$, $\delta = +62\arcdeg37\arcmin52\farcs3$ (J2000) and $\alpha = 22^{\rm{h}}56^{\rm{m}}20\fs3$, $\delta = +62\arcdeg41\arcmin43\farcs0$ (J2000).  The orientation of the fields long axis is $P.A.=36.3\arcdeg$. These fields were imaged with a $3\times4$ mosaic of adjacent positions separated by $290\arcsec$, and images were taken at 12 dithered positions for each of the 12 cells of the mosaic. The images were taken in the high-dynamic-range mode with long effective exposures of 96.8~s in 3.6, 4.5, 5.8~$\mu$m channels, and two 46.8~s exposures in the 8.0~$\mu$m channel, as well as short effective exposures of 0.4~s in all 4 channels. Full coverage in all four channels is obtained for a $\sim 13\arcmin \times 15\arcmin$ area, which in turn covers $\sim 65\%$ of the $17\arcmin \times 17\arcmin$ {\it Chandra's} ACIS-I field, as shown in Figures \ref {fig_map} and \ref{fig_color_image}.

Basic Calibrated Data (BCD) products from the Spitzer Science Center's IRAC pipeline version S15.3.0 were automatically treated with the WCSmosaic IDL package developed by R. Gutermuth for the IRAC instrumental team. Starting with BCD data products, the package mosaics individual exposures while treating bright source artifacts, cosmic ray rejection, distortion correction, subpixel offsetting, and background matching \citep{Gutermuth08}. We selected a plate scale of 0.86$\arcsec$ for the reduced IRAC mosaics, which is the native scale divided by $\sqrt{2}$.

Aperture photometry of IRAC sources was obtained using the IRAF task PHOT. For the most of the X-ray and non-Xray (\S \ref{non_xray_subsection}) selected sources, an aperture radius of 4 pixels (3.44\arcsec) was used with an adjoining sky annulus width of 1 and 6 pixels (0.86\arcsec and 5.16\arcsec) for the 3.6/4.5 and 5.8/8.0~$\mu$m bands, respectively. For the crowded sources, aperture radii of 2 or 3 pixels (1.72\arcsec and 2.58\arcsec) and adjoining sky annulus width of 1 pixel (0.86\arcsec) in all 4 bands were used. We adopted zero point magnitudes ($ZP$) of 19.670, 18.921, 16.855, and 17.394 in the 3.6, 4.5, 5.8, and 8.0~$\mu$m bands, where $M = -2.5 \log (DN/sec) + ZP$ \citep{Reach05}. The total aperture corrections applied to our measurements are: 0.185, 0.175, 0.165, 0.250 in the case of 4 pixel aperture; 0.275, 0.305, 0.515, 0.735 in the case of 3 pixel aperture; and 0.655, 0.735, 1.005, 1.010 in the case of 2 pixel aperture in the 3.6, 4.5, 5.8, and 8.0~$\mu$m bands, respectively. The reported photometric errors include Poisson errors in the source and background emission plus a 2\% uncertainty in the calibration of IRAC \citep{Reach05}.

\subsection{IRAC Counterparts to X-ray Sources \label{cross_identification_subsection}}

The initial list of point sources in these images was produced by running IRAF task STARFIND with a low detection threshold in order to identify as many sources as possible, permitting false detections in this stage of the analysis. An automated cross-correlation between the {\it Chandra} source positions of \citet{Getman06} and IRAC candidate source positions was made using a search radius of $2\arcsec$ within $\sim 6\arcmin$ of the ACIS field center, and a search radius of $3.5\arcsec$ in the outer regions of the ACIS field where X-ray source positions are more uncertain due to the deterioration of the {\it Chandra} telescope point spread function (Figure \ref{fig_color_image}). This was followed by a careful visual examination of each source in both bands to remove dubious sources and associations.  

Of the 431 X-ray sources within the $17\arcmin \times 17\arcmin$ {\it Chandra} ACIS-I field \citep{Getman06}, 396 (92\%) lie within the 3.6/5.8~$\mu$m IRAC coverage and 412 (96\%) lie within the 4.5/8.0~$\mu$m IRAC coverage.  Out of [396, 412, 396, 412] X-ray sources in the [3.6, 4.5, 5.8, 8.0~$\mu$m] channels, [384, 399, 369, 385] are identified with IRAC sources. The IRAC detection rate of X-ray sources is thus $\sim 97\%$ for 3.6 and 4.5~$\mu$m channels and $\sim 93\%$ for 5.8 and 8.0~$\mu$m channels.

2MASS associations are obtained as described in \citet{Getman06}. Optical counterparts were identified from the $VI_c$ photometric catalog of \citet{Mayne07}\footnote{The optical survey encompasses most of the Chandra-ACIS-I field and extends further from the ACIS-I $\sim 15\arcmin$ to the north-west and $\sim 20\arcmin$ to the west.}.  Unlike the search for the IR counterparts, we lacked the ability to visually examine the $Chandra$-optical matches. We therefore compared the $Chandra$-optical offset (Off$_1$ in Table~\ref{tbl1})  with the $Chandra$-$2MASS$ and $Chandra$-IRAC positional offsets (Off$_2$ and Off$_3$).  Accepted optical matches typically have $|\rm{Off}_1 - \rm{Off}_{2,3}| \la 0.2\arcsec$. Out of the total of 431 $Chandra$ X-ray sources, 368 are in the field of view of the optical survey and 321 of them have optical counterparts.  The detection rate of the X-ray sources in the optical survey is thus $\sim 87\%$.  

There are [12, 13, 27, 27] X-ray sources in the [3.6, 4.5, 5.8, 8.0~$\mu$m] channels lacking mid-IR counterparts.  Roughly half of these lie in the interface region between the Cep~B molecular cloud and S~155 H{\sc II} region where the IRAC point source sensitivity is reduced by the high background nebular emission from heated dust (Figure \ref{fig_map}).  These are probably true cluster members, but are omitted from the science analysis below.  The other half of these unidentified X-ray sources is uniformly distributed outside of the Cep~B cloud with weak X-ray fluxes and high X-ray median energies consistent with extragalactic background sources.  Most of these are listed in Table 4 of \citet{Getman06} as likely extragalactic contaminants.  

The results of this search for stellar counterparts of the $Chandra$ sources are presented in Table~\ref{tbl1}.  Listed by $Chandra$ sequence number, the table gives offsets between the optical/IR stars and the X-ray sources, $VI_cJHK_s$ and IRAC photometry, and photometric flags.  The final column gives the aperture used in the photometry for each IRAC band; a `9' value indicates that the X-ray source lies outside the IRAC coverage in that band or the IRAC source is too weak or strongly contaminated with no reliable photometry.

\section{PRE-MAIN SEQUENCE DISK CLASSIFICATION \label{yso_classification_section}}  

We base the evolutionary classification of X-ray emitting PMS stars in the Cepheus region on a comparison of their IR spectral energy distributions (SEDs) with the SEDs of PMS stars in the well-studied IC~348 cluster in the Perseus molecular cloud \citep{Lada06}. IC~348 stars have roughly the same  $2-3$~Myr age as Cep~OB3b stars.  The categories are: Class~I (protostar, disk and envelope), Class~II (PMS star and accretion disk), and Class~III (PMS star with weak or absent disks).  A simple characterization of IR SEDs involves the SED spectral slope $\alpha$ of the SED $\alpha = d \log(\lambda F_{\lambda})/d \log(\lambda)$ over the range $2 < \lambda < 20 \mu$m. Class~I sources have $0 < \alpha \la 3$, Class~II sources have $-2 \la \alpha \leq 0$, and Class~III sources have $-3 < \alpha \la -2$. Using optical and IR data for the IC~348 PMS stars, \citet{Lada06} showed the utility of the SED slope measured in the IRAC wavelength range from 3.6 to 8.0~$\mu$m.  Comparing optical and IRAC SEDs to disk models calculated from a Monte Carlo radiative transfer code \citep{Wood02,Walker04}, they find that Class~II systems with optically thick disks have dereddened IRAC slopes $\alpha_{d} > -1.8$, ``transition disk'' (TD) systems with inner disk holes or optically thin inner disks have $ -2.56 < \alpha_{d} < -1.8$, and diskless Class~III systems have $\alpha_{d} < -2.56$.  \citet{Lada06} also provide useful empirical templates of median SEDs of IC~348 PMS stars for each evolutionary class.

We proceed with the disk classifications through the comparison of the observed Cepheus SEDs in $2MASS+$IRAC IR bands to the (de)reddened median SED templates of IC~348 PMS stellar photospheres. We further confirm these classifications through calculations of the SED slope $\alpha$,  and locations in IRAC color-color diagrams. Results of our disk classification are shown in Table~\ref{tbl2}.

\subsection{Comparison with IC~348 \label{IC348_subsection}}

First, a rough estimate of individual stellar masses of the Cepheus PMS stars is made by de-reddening star positions in the $J$ versus $J-H$ color-magnitude diagram to the $3$~Myr and $2$~Myr theoretical isochrons for the Cep~OB3b/S~155 and Cep~B subregions, respectively, based on the models of \citet{Siess00} and \citet[][see Appendix A]{Baraffe98}. These individual stellar photometric mass and extinction estimates are listed in Table \ref{tbl2}. Although individual mass estimates may be subject to significant uncertainties and may not be always compatible with masses obtained by more accurate methods such as optical spectroscopy, this approach should be adequate to assign an IC~348 SED template \citep[Table~4 of][]{Lada06} to individual Cepheus PMS stars.

Second, it is important to note that the IC~348 template SEDs are observed (not de-reddened) and exhibit an extinction of $A_V \sim 2-2.5$~mag \citep{Lada06}. Thus for each of the Cepheus stars, an IC~348 template SED at the corresponding spectral sub-class is first reddened and de-reddened within the wide but finite range of extinctions using reddening relationships described in Appendix~A for $JHK_s$ and \citet{Flaherty07} for IRAC bands. The (de)reddened IC~348 template which, after normalization to an observed $J$-band Cepheus SED point, is the closest to an observed $H$-band Cepheus point (both $J$ and $H$-band points are considered to be indicators of a pure photospheric emission) is recognized here as the best-fit to Cepheus data. Third, the best-fit IC~348 (de)reddened median template is visually matched to the observed Cepheus SED, allowing classification of the Cepheus stars. 

Out of the total of 431 X-ray Cepheus sources, our classification procedure yields 215 diskless Class~III systems  (labelled ``NoD'' in Table~\ref{tbl2}) and 139 Class~II or I disk-bearing systems (``DSK'').  Based on the membership analysis of \citet{Getman06}, 24 X-ray sources without IR counterparts are possible extragalactic contaminants (``EXG'') and 13 are foreground candidates (``FRG'').  Forty Chandra sources with IR detections which are unreliable or insufficient for classification have an uncertain classification (``UNC'').   We further subjectively classify 20 stars as Class~II/III transition disks (labelled as ``TD'' and ``DSK''); these follow the photospheric SED at wavelengths shorter than at least $\sim 3.6\mu$m. Four $Chandra$ sources ( \#\# 314, 322, 328, and 390) embedded in the Cep~B core have extreme SED excesses with $\alpha > 0$ and are classified as Class~I stars. 
 
\begin{figure}
\centering
\includegraphics[angle=0.,width=7.0in]{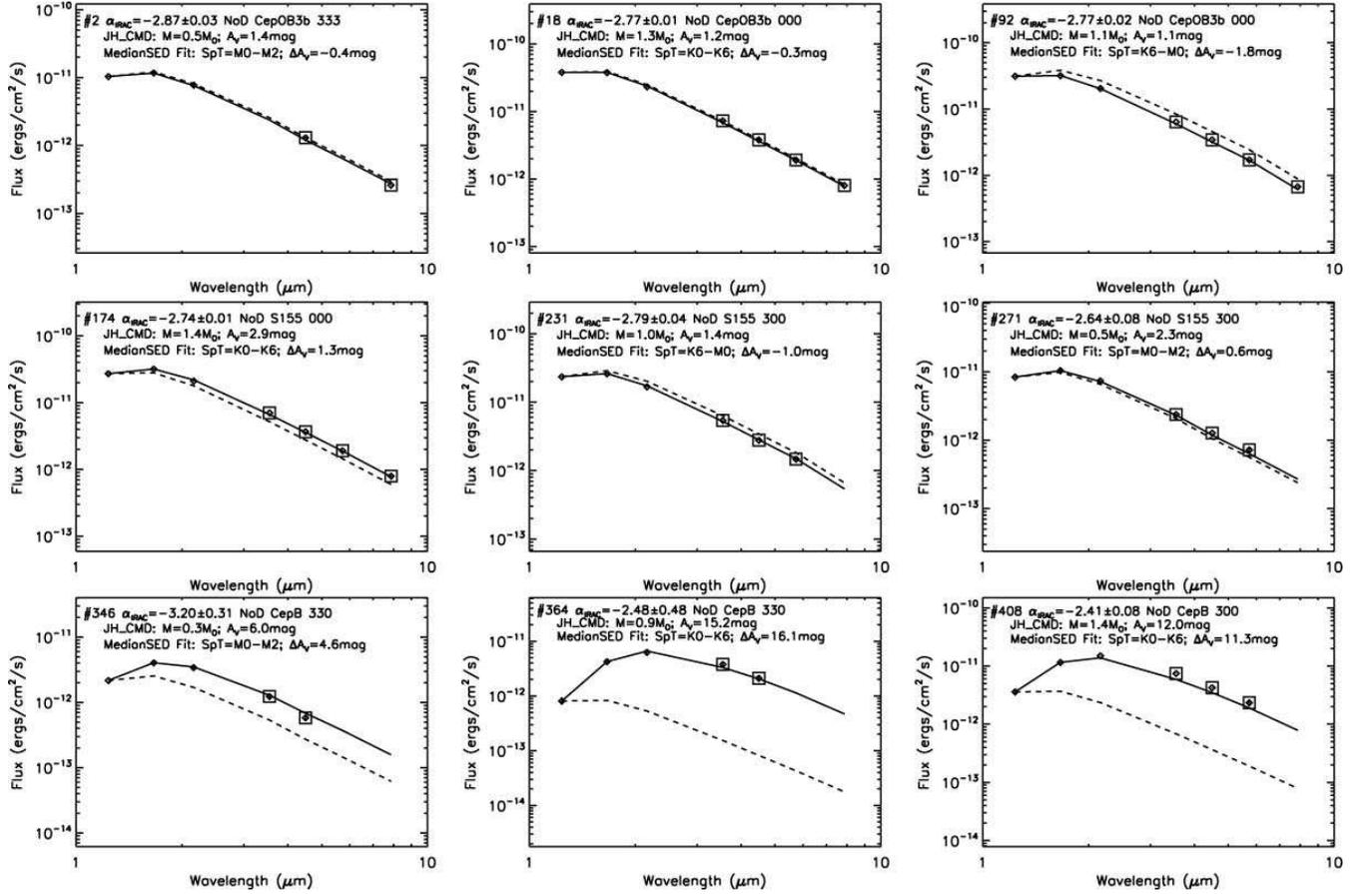}
\caption{IR spectral energy distributions for nine representative diskfree (Class~III) X-ray stars. From top to bottom, SEDs for stars in Cep~OB3b, S~155 and Cep~B subregions. $JHK_s$ (diamond) and IRAC-band (square) flux points with usually small errors. The dashed and solid lines give the original and (de)reddened IC~348 median SED from \citet{Lada06} fitted to the data. The top two lines of the panel labels give information from Table~\ref{tbl2}. The third line gives the spectral class of the IC~348 median SED template from \citet{Lada06} and reddening applied to the original template SED to fit the observed Cepheus source SED.  See the electronic edition of the Journal for the full atlas of IR SEDs of the Cepheus X-ray young stars. \label{fig_sed_nod}}
\end{figure}

\begin{figure}
\centering
\includegraphics[angle=0.,width=7.0in]{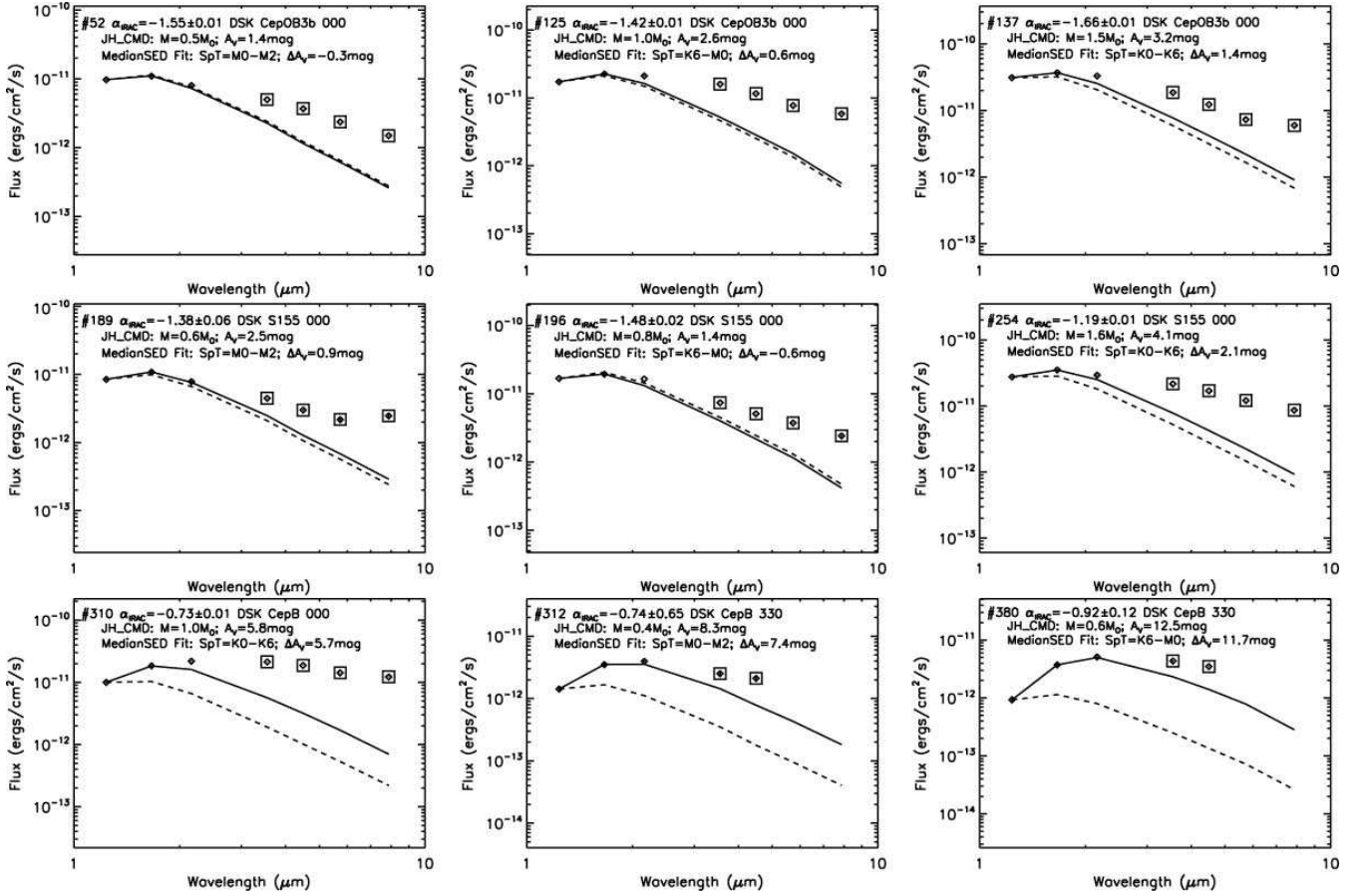}
\caption{IR spectral energy distributions for nine representative disk-bearing (Class~II) X-ray stars.  See Figure~\ref{fig_sed_nod} for details. \label{fig_sed_dsk}}
\end{figure}

\begin{figure}
\centering
\includegraphics[angle=0.,width=7.0in]{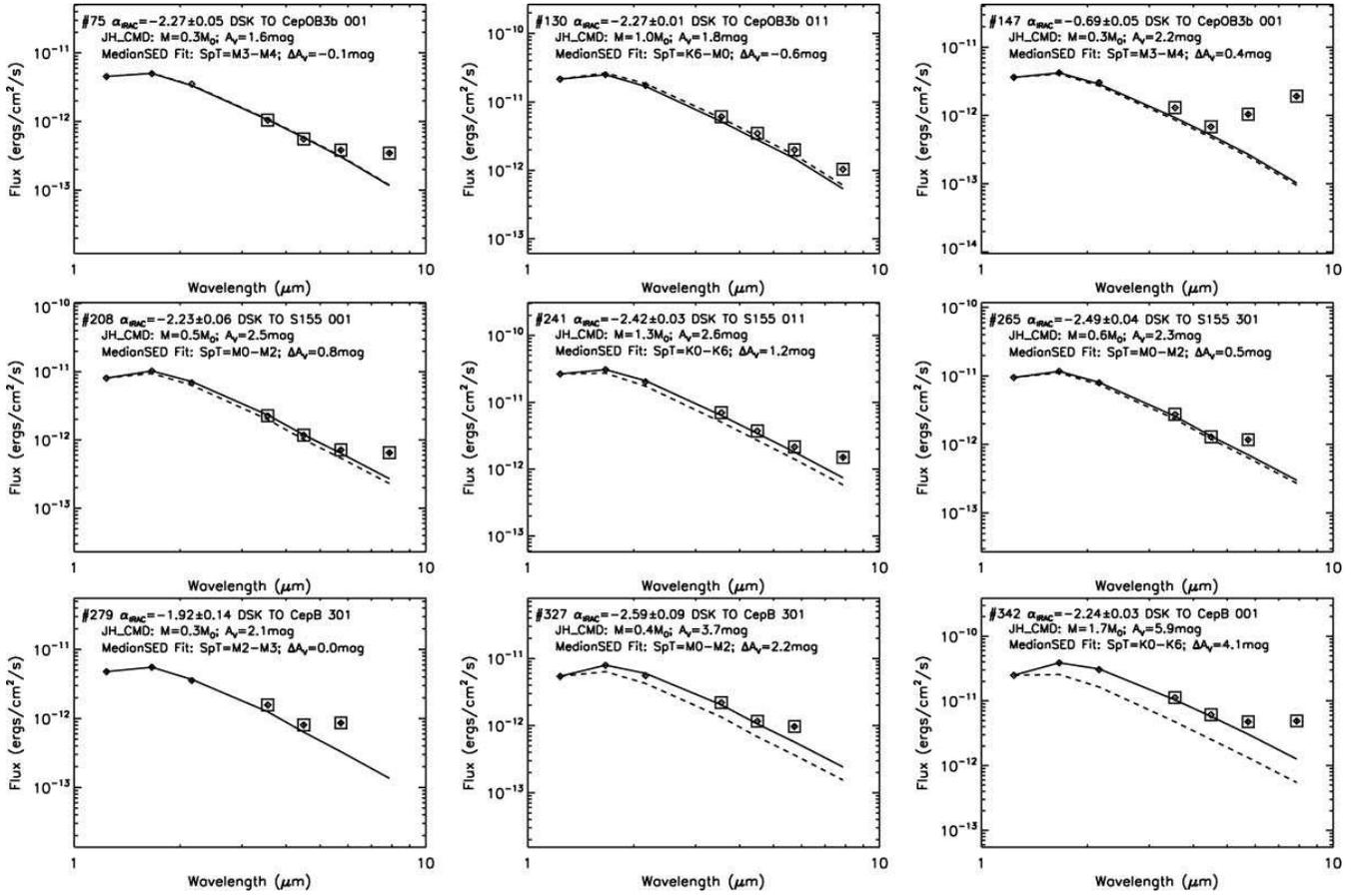}
\caption{IR spectral energy distributions for nine representative transition disk (Class~II/III) X-ray stars. See Figure~\ref{fig_sed_nod} for details.  \label{fig_sed_tos}}
\end{figure}

Examples of diskfree, disk-bearing, and transition SEDs are shown in Figures~\ref{fig_sed_nod}-\ref{fig_sed_tos}. The complete set of SEDs for 354 X-ray emitting Cepheus PMS stars classified here is available on-line as a Figure Set.  The on-line atlas contains 161 diskfree stars of the Cep~OB3b region (labelled as ``NoD'' and ``CepOB3b'' in Table~\ref{tbl2}); 75 disk-bearing stars of the Cep~OB3b (``DSK'' and ``CepOB3b'');  36 diskfree stars of the S~155 region (``NoD'' and ``S~155''); 26 disk-bearing stars of the S~155 region (``DSK'' and ``S155''); 18 diskfree stars in the Cep~B region (``NoD'' and ``CepB''); and 38 disk-bearing stars in the Cep~B region (``DSK'' and ``CepB'').  Quantitative information derived from the fitting process is given in Table~\ref{tbl2} including the masses and absorptions estimated from the $J-(J-H)$ color-magnitude diagram (columns 4 and 5) and the SED classification obtained by comparison with IC~348 templates (columns 6 and 7).

\subsection{Calculation of SED slope $\alpha$ \label{slope_subsection}}

\begin{figure}
\centering
\includegraphics[angle=0.,width=6.5in]{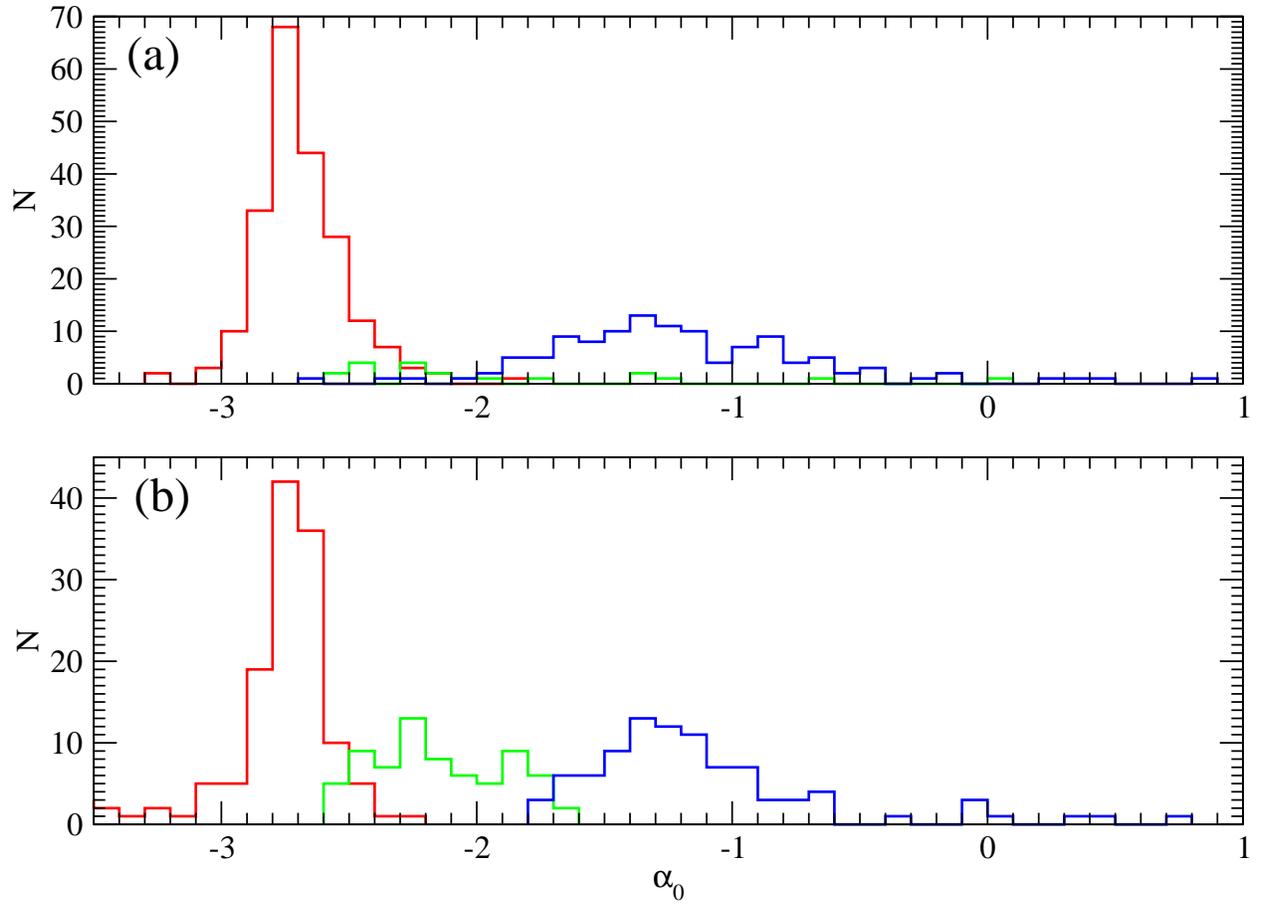}
\caption{Histograms of observed SED spectral slopes $\alpha_0$ for (a) X-ray stars in the Cepheus field and (b) comparisons in IC~348. Colors indicate PMS evolutionary classes: diskfree Class~III stars (red), disk-bearing Class~I and II stars (blue), and transition disk Class~II/III stars (green). \label{fig_slope_histogram}}
\end{figure}

We performed least-squares linear fits to the $\log(\lambda F_{\lambda})$ values in the four IRAC wavelength band to obtain the observed (not de-reddened) SED spectral index, $\alpha_{0}$.  These are tabulated in column~2 of Table~\ref{tbl2}. Figure~\ref{fig_slope_histogram} shows the histogram of 354 $\alpha_{0}$ values for the X-ray selected Cepheus PMS stars compared to 299 IC~348 PMS stars from \citet{Lada06}.

Two important features are noted. First, the Cepheus histogram shows a bimodal distribution with peaks at $\alpha_{0} \sim -1.3$ and $-2.7$ corresponding to disk-bearing and diskless star samples, respectively.  The IC~348 histogram has the same peaks, and similar bi-modality is seen in the PMS populations of the Cha~I cloud \citep{Luhman08a} and $\sigma$~Ori cluster \citep{Hernandez07}.  Second, in contrast to the IC~348 population, the Cepheus region does not produce a rich TD population in the interval $-2.6 < \alpha_0 < -1.8$.  The X-ray selected Cepheus population has 20/139 ($14\%$) TDs among disk-bearing stars compared to 70/163 ($43\%$) in IC~348.  This apparent difference in TD frequency is discussed in \S~\ref{to_subsection}.

\subsection{IRAC Color-Color Diagrams \label{ccd_subsection}}
   
\begin{figure}
\centering
\includegraphics[angle=0.,width=7.0in]{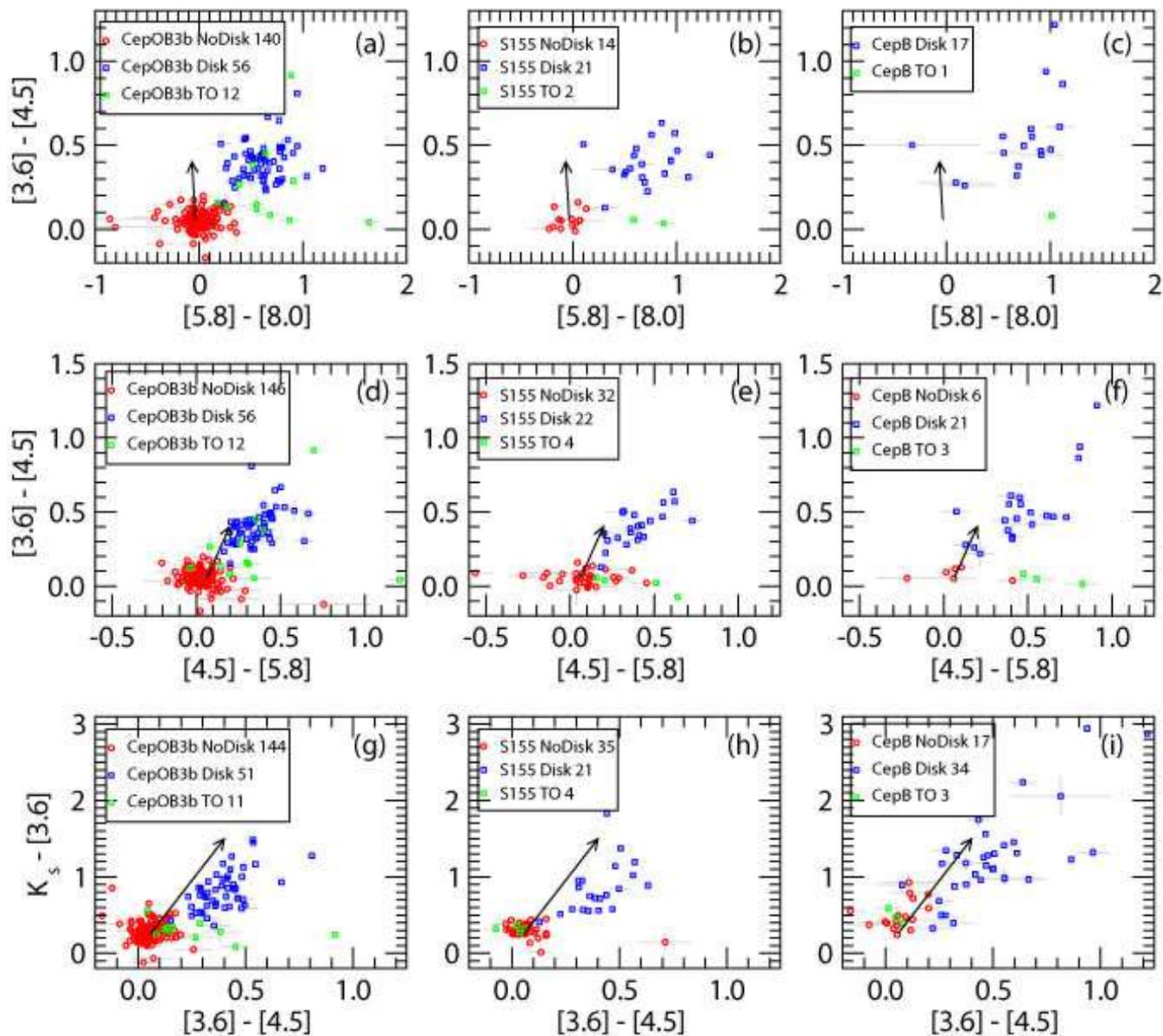}
\caption{Infrared color-color diagrams of X-ray stars:  Cep~OB3b (left), S~155 (middle) and Cep~B (right) subregions; long IRAC channels (top), short IRAC channels (middle), 2MASS $K_s$ and short IRAC channels (bottom).  Colors indicate PMS evolutionary classes: Class~III (red), Class~I and II (blue), and Class~II/III transition disk stars (green).  The reddening vectors show $A_V \sim 30$~mag using the extinction law from \citet{Flaherty07}. Legends indicate number of stars with available photometry. \label{fig_ccds}}
\end{figure}

Comparison of $Spitzer$-IRAC source positions in the $[3.6] - [4.5]$ $vs.$ $[5.8] - [8.0]$ diagram with the results of models of stars with dusty disks and envelopes led \citet{Allen04} and \citet{Megeath04} to propose this diagram as an excellent tool for PMS classification.  \citet{Hartmann05} show that similarly successful PMS classification schemes can be obtained at shorter wavelengths using the $[3.6] - [4.5]$ $vs.$ $[4.5] - [5.8]$ and $K_s - [3.6]$ $vs.$ $[3.6] - [4.5]$ diagrams. In Figure \ref{fig_ccds}, we compare the SED-based classifications of Cepheus X-ray stars derived in \S \ref{IC348_subsection} with the expected loci of PMS stars in IRAC color-color diagrams.  Here we also begin comparison of the disk characteristics in three spatial subregions: the unobscured Cep~OB3b cluster, the optically-bright S~155 nebula, and the Cep~B molecular cloud core.  The regions are measured radially from the location of the hot molecular core at $\alpha = 22^{\rm{h}}57^{\rm{m}}15\fs2$, $\delta = +62\arcdeg37\arcmin11\farcs1$ (J2000).  

Figure \ref{fig_ccds} shows the three IRAC color-color diagrams for the sources in the three subregions.  Seventy-seven Cep~OB3b sources are missing from the plots due to inadequate photometry, often due to location outside of the full four-band coverage.   Sources are also omitted from the S~155 and Cep~B areas when photometry at the longer 5.8~$\mu$m and 8.0~$\mu$m bands are absent or inaccurate due to bright nebular emission. The positions of stars in the color-color diagrams generally agree with our previous classification of stars as disk-bearing and diskless PMS stars based on comparison with IC~348 stars (\S \ref{IC348_subsection}), shown as blue and red symbols in Figure~\ref{fig_ccds}. The $F_2$ flag of Table~\ref{tbl2} indicates sources with their disk classification from \ref{IC348_subsection} inconsistent with the simple color criterion that diskless stars have  $[3.6] - [4.5] \la 0.2$, $[4.5] - [5.8] \la 0.2$ or $[5.8] - [8.0] \la 0.2$. Several sources with such ``discrepant'' locations on color-color diagrams may represent additional cases of transitional disks.

\section{NON-CHANDRA INFRARED-EXCESS MEMBERS \label{non_xray_subsection}}

Due to the relatively short $Chandra$ exposure and large distance to the Cepheus region, the X-ray PMS sample is only complete down to $\sim 0.5$~M$_{\odot}$ for Cep~OB3b/S~155 Class~III stars, less complete for Cep~OB3b/S~155 Class~II stars, and even less complete for Cep~B embedded objects (\S \ref{IMF_subsection}). As the stellar Initial Mass Function peaks around $\sim 0.3$~M$_\odot$, hundreds of Cepheus stars should be present in the $Chandra$ field of view with their X-ray luminosities below the sensitivity limits of the $Chandra$ observation of $\log L_x \sim 29-29.5$ erg~s$^{-1}$ \citep{Getman06}.  With an age around $\sim 2-3$~Myr and younger (Appendix~A), a large fraction of the Cepheus stars with masses below 0.5~M$_\odot$ in our field will have optically thick disks detectable with IRAC.  Selection by IRAC IR-excess should even detect many brown dwarfs in the region.  Shallow IRAC exposures efficiently detected disks around brown dwarfs in the Orion cloud at a distance of $\sim 400$~pc over several magnitudes in brightness \citep[][and references therein]{Luhman08b}.  The sensitivity at the Cepheus distance around $\sim 725$~pc will be reduced by $\sim 1.3$~mag, sufficient to detect a considerable fraction of the disk-bearing brown dwarfs. Thus some fraction of non-X-ray PMS stars with lower-masses are expected to be detected in the IRAC survey within the $Chandra$ field of view.

\begin{figure}
\centering
\includegraphics[angle=0.,width=7.0in]{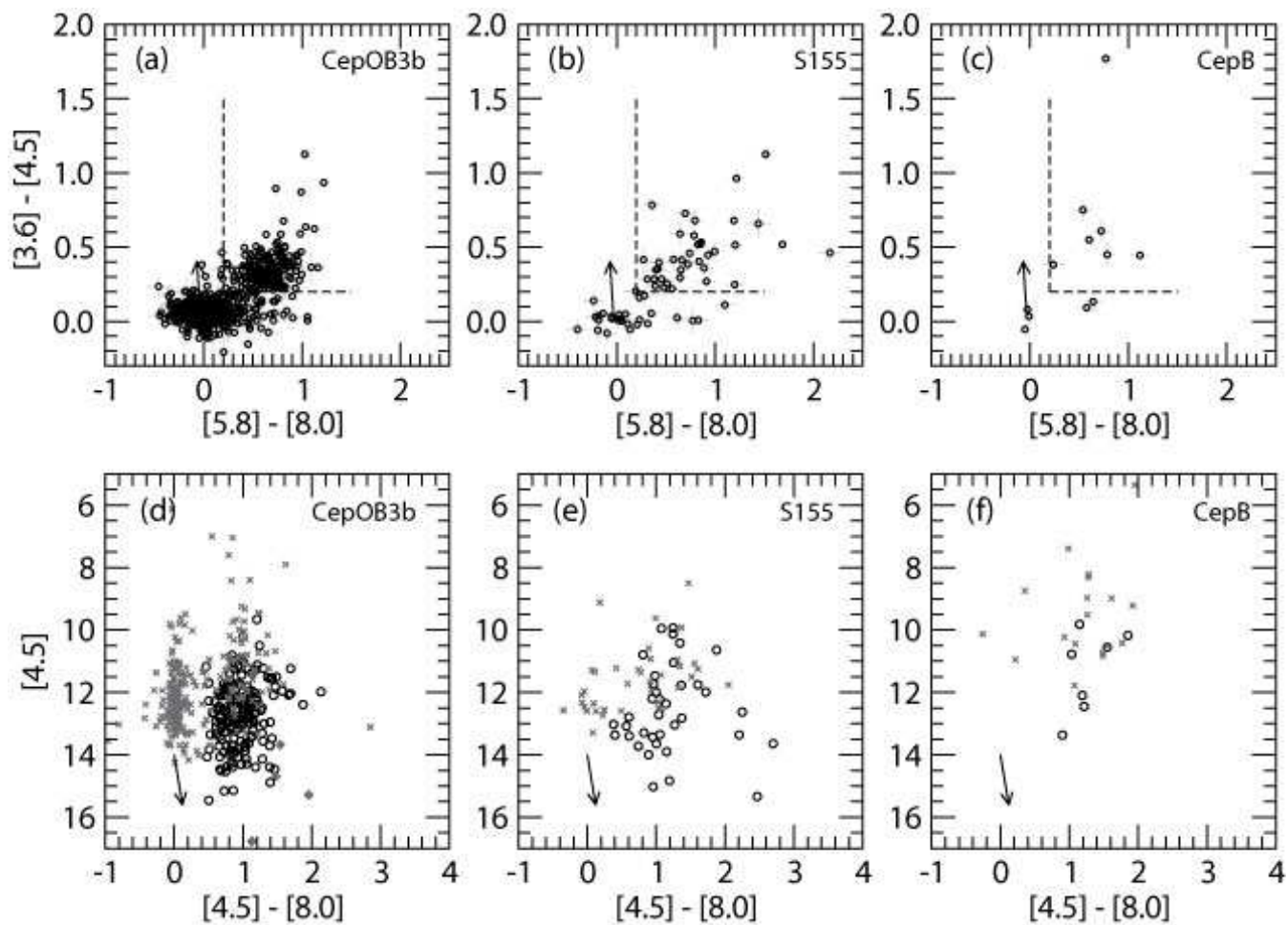}
\caption{IRAC color-color diagrams of non-X-ray IR-excess sources.  Top: $[3.6]-[4.5]$ versus $[5.8]-[8.0]$ color-color diagrams employed to select PMS candidates for (a) Cep~OB3b, (b) S~155, and (c) Cep~B subregions.  Dashed lines show classification criteria.  Bottom:  $[4.5]$ versus $[4.5]-[8.0]$ color-magnitude diagrams for (d) Cep~OB3b, (e) S~155, and (f) Cep~B subregions showing X-ray emitting PMS stars ($\times$) and non-X-ray PMS candidates (circles).  Panel (d) also shows X-ray emitting extragalactic contaminants (diamonds).  Reddening vectors represent $A_V \sim 30$~mag. \label{fig_ccd_cmd_nonxray}}
\end{figure}

Figure~\ref{fig_ccd_cmd_nonxray}$a-c$ shows IRAC color-color diagrams for 774 IRAC sources in the $Chandra$ ACIS-I field of view that are not associated with X-ray sources.  We restrict these diagrams to sources with photometric errors $<0.1$~mag in all four IRAC bands. Using the IR color-color classification criteria presented in \S~\ref{ccd_subsection}, we identify 224 non-X-ray IR-excess stars.  Table~\ref{tbl3} gives IR properties for the 224 non-X-ray IR-excess sources.  

Figure~\ref{fig_ccd_cmd_nonxray}$d-f$ compares X-ray-selected and IR-excess-selected samples in the $[4.5]$ $vs.$ $[4.5]-[8.0]$ color-magnitude diagram. As expected, the majority of these new PMS candidates are fainter than the X-ray selected sample, although there is a considerable overlap in the distributions around $14<[4.5]<12$~mag ($\sim 0.1-0.3$~M$_{\odot}$).  Most of these new stars thus have masses in the $<0.5$~M$_\odot$ and some in sub-stellar range. Panels  show that, despite the different selection criteria, the color distributions of these IRAC-selected stars are similar to those of the X-ray selected PMS stars.   However, several of the weakest and reddest ($[4.5] \ga 14$~mag and $[4.5]-[8.0] \ga 0.5$~mag) IR-selected sources may be unrelated extragalactic objects  \citep{Harvey06}.  Four probable extragalactic contaminants from the X-ray-selected sample lie near this region of the diagram (grey diamonds in Figure \ref{fig_ccd_cmd_nonxray}$d$).

In the unobscured Cep~OB3b subregion, the combined X-ray-selected and IR-excess-selected sample should give a complete census of the disk-bearing population down to $0.2-0.3$~M$_{\odot}$, a nearly complete census of the disk-free population down to $0.5$~M$_\odot$, and a fraction of both the disk-bearing and disk-free populations into the brown dwarf regime (\S \ref{IMF_subsection}).  But due to the limited point source sensitivity from the non-uniform IR nebular emission, the identified PMS populations of the S~155 and Cep~B subregions are less complete.  This can be seen from the paucity of sources fainter than $[4.5]>14$~mag in S~155 and $[4.5]>13$~mag in Cep~B  (Figure \ref{fig_ccd_cmd_nonxray}$e-f$). More accurate information on mass completeness limits of Cepheus PMS stars is given in the next section.

\section{INITIAL MASS FUNCTION OF THE CEPHEUS CLUSTER \label{IMF_subsection}}

In the area covered by the $Chandra$ field and all four bands of the IRAC mosaic, we examine the stellar IMF separately for diskless and disk-bearing stars in each of the three sub-regions. For a given mass, the X-ray detection efficiency of Class~III stars is somewhat high than that of Class~II stars (Appendix \ref{lx_disks_section}). To compensate for this effect, in this IMF and the following disk evolution analyses, we use the disk-bearing stellar sample that is the combination of the $Chandra$ and non-$Chandra$ IR-excess member young stars (\S \ref{non_xray_subsection}) while retaining the diskless stellar sample as purely composed of X-ray stars. In this IMF analysis section, we exclude X-ray stars lying outside of the 4 IRAC band mosaic, as well as the most massive young star in the region, O7Vn star HD~217086 ($Chandra$ \# 240) which is the only star with mass above $4$~M$_{\odot}$ (upper boundary of considered IMF mass range; Figure \ref{fig_imf}). We also exclude 51 non-$Chandra$ IR-excess possible members with inferred masses below 0.1, as some of them may still be extragalactic background objects (\S \ref{non_xray_subsection}).

Figure \ref{fig_imf}$a$ shows that the mass distributions of the Cep~OB3b sub-region stars, both diskless (red) and disk-bearing (blue), nicely follow the shape of the Galactic field IMF, showing a Salpeter powerlaw slope between $0.5-3$~M$_\odot$, and a peak around $0.3-0.5$~M$_{\odot}$.  The decline seen below $0.3-0.5$~M$_\odot$ is attributable to our incompleteness limits and is probably not intrinsic to the cluster. The pure $Chandra$ Class~III sample is complete down to $0.5$~M$_{\odot}$, while the combined $Chandra$ and non-$Chandra$ Class~II sample is complete for $\ga 0.3$~M$_{\odot}$ Cep~OB3b stars. Assuming a Galactic field IMF shape, we estimate the total PMS population in the Cep~OB3b sub-region to be $\sim 750$ stars down to 0.1~M$_{\odot}$.

An important negative result here is that we do not confirm the tentative explanation by \citet{Getman06} that the non-standard X-ray luminosity function of the Cep~OB3b is due to an anomalous `bottom-heavy' IMF with an excess of $0.3$~M$_{\odot}$ stars. Although a slight difference between the combined $Chandra$ and $Spitzer$ sample and the Galactic field IMF is seen (Figure \ref{fig_imf}$a$), it is not statistically significant. 

Figure \ref{fig_imf}$b$ shows that mass distributions of the $S~155$ sub-region stars, although with fewer stars than in Cep~OB3b, also follow the Galactic field shape with mass completeness limits of $0.5$~M$_{\odot}$ for both Class~II and Class~III stellar samples with an estimated total PMS population of $\sim 200$ S~155 stars down to 0.1~M$_{\odot}$.

The observed mass distribution of the embedded Cep~B cluster may be either top-heavy or, more likely, simply lacking unidentified low-mass disk-free PMS stars due to $Chandra$ sensitivity loss from obscuration (Figure \ref{fig_imf}$c$). A similar situation is noticed in the study of the embedded population of the bright-rimmed cloud IC~1396N \citep{Getman07}. In the case of usual IMF with still unidentified low-mass stars, the observed Cep~B stellar population may be complete only somewhere above $1$~M$_{\odot}$, and the total intrinsic PMS population of the Cep~B may reach a few hundred stars. The triggered population is thus much larger than that identified by early radio and infrared techniques \citep{Testi95}.

\section{DISK EVOLUTION \label{results_section}}
  
\subsection{Mass Dependence \label{mass_domain_subsection}}

Growing evidence has emerged from IRAC young cluster studies that more massive PMS stars have shorter disk lifetimes. \citet{Luhman08a,Luhman08b} find that disk fractions in the $2-3$~Myr old IC~348 and $\sigma$~Ori clusters decrease from $\sim 50-60\%$ in brown dwarfs to $\sim 30-40\%$ in stars with $M \ga 1$~M$_{\odot}$. In the older 5~Myr old Upper Sco and NGC~2362 clusters, the disk fraction decreases from $10-20\%$ in $0.1-1$~M$_{\odot}$ stars to $\la 1\%$ in higher-mass stars \citep{Carpenter06, Dahm07}.  The trend may not be universal: the disk fraction of $M \ga 1$~M$_{\odot}$ stars in the $2$~Myr old Cha~I cluster is $> 60\%$ \citep{Luhman08a}, much higher than that found in IC~348 and $\sigma$~Ori at similar ages.

The Figure \ref{fig_mj_diskfraction} compares the mass-dependent disk fractions of Cep~OB3b and the other three regions examined by \citet{Luhman08b}.  Our sample here is the combined 215 $Chandra$ (classified as ``DSK'' or ``NoD'' in Table \ref{tbl2}) and 142 disk-bearing non-$Chandra$ (Table \ref{tbl3}) Cep~OB3b sources within the overlap area of 4 IRAC band and $Chandra$ fields, omitting the most massive young star in the region, O7Vn star ($Chandra$ \# 240) with $M_J < 0$~mag, and 37 non-$Chandra$ IR-excess possible members of Cep~OB3b with inferred masses below $0.1$~M$_{\odot}$. Appendix \ref{lx_disks_section} shows that the relatively short $Chandra$ observation is only sensitive to stars down to $0.2$~M$_{\odot}$, so we have no Cep~OB3b coverage for very low mass Class~III stars and brown dwarfs. Cep~OB3b stars have average ages around $2-3$~Myr old (Appendix \ref{distance_age_subsection}) similar to the ages in Cha~I, IC~348, and $\sigma$~Ori; possibly younger stars in the S~155 and Cep~B region are omitted here. Following Luhman et al., the absolute $J$-band magnitude $M_J$ serves as a proxy of mass.  Our $M_J$ values are obtained using individual $A_J$ extinctions from Tables~\ref{tbl2}-\ref{tbl3}, which in turn were derived through the evolutionary model-dependent procedure assuming the distance of $725$~pc and the age of $3$~Myr.  Individual magnitude accuracies should be sufficient to reliably assign individual sources to the coarse, 2~mag-wide $M_J$ bins.  Errors on disk fractions have been estimated using binomial distribution statistics as described by \citet{Burgasser03}. For the Cep~OB3b stars, $M_J$ bins have been systematically shifted by $+0.3$~mag from those of other regions in order to include in the second $M_J$ bin the $\sim 0.5$~M$_{\odot}$ mass stars, stars for which the combined $Chandra$ and non-$Chandra$ Cep~OB3b sample is complete (\S \ref{IMF_subsection}). 

Figure \ref{fig_mj_diskfraction} shows that the disk fraction of the Cep~OB3b stars (circles) for the two highest mass bins is around 45\% and agrees with the fractions seen in IC~348, but appears higher than the comparison clusters for the lowest $< 0.5$~M$_\odot$ mass bin. The most plausible explanation for the reduced disk fraction in the $<0.5$~M$_\odot$ bin is incompleteness of the Cep~OB3b sample at low masses (\S \ref{IMF_subsection}).  Here, our short $Chandra$ observation is not sensitive to many low-mass Class~III stars and the observational bias towards non-$Chandra$ disk-bearing stars affects the sample selection.

We thus find no significant dependence of disk fraction on mass in the Cep~OB3b $Chandra$ plus non-$Chandra$ selected sample above $0.5$~M$_{\odot}$\footnote{The results are similar with a choice of longer distance of $\sim 850$~pc and thus younger age of $\sim 2$~Myr for the Cep~OB3b cluster (Appendix~\ref{distance_age_subsection}).}. This agrees with the lack of mass dependence seen in IC~348 and $\sigma$~Ori in the $>0.2$~M$_{\odot}$ regime.  The increased fractions reported in these clusters were found in the $<0.2$~M$_{\odot}$ regime which is not covered by our Cep~OB3b selected sample.  However, we do not support studies in other clusters that report a rapid loss of IRAC-band disk emission in intermediate-mass stars.  These studies might consider possible sample incompleteness effects, such as an under-sampling of diskfree Class~III low-mass members.  Our inclusion of the X-ray selected sample is specifically designed to be nearly free of this bias.

\subsection{Spatial Gradients \label{spatial_domain_subsection}}

We now consider spatial distribution of the disks around the mass-complete $Chandra$ plus non-$Chandra$ PMS samples over the area covered by the 4 IRAC mosaics and the $Chandra$ field. Figure \ref{fig_spat_distrib} shows the spatial distribution, and Table \ref{tbl_frac} provides quantitative details for two mass strata, two azimuthal zones (NE and SW), three radial subregions (Cep~OB3b, S~155, and Cep~B), and three radial layers discussed below (inner, intermediate, and outer).  The azimuthal zones are separated by the line that roughly bisects the optically-bright (Figure \ref{fig_map}) and IR-bright nebular emission. Radial distances are measured from the hot core of the molecular cloud ($\alpha = 22^{\rm{h}}57^{\rm{m}}15\fs2$, $\delta = +62\arcdeg37\arcmin11\farcs1$ (J2000)). Below we consider the Cepheus sample of 220 $Chandra$ plus non-$Chandra$ PMS stars with stellar masses above $0.5$~M$_{\odot}$, the mass completeness limit for Cep~OB3b/S~155 stars (\S \ref{IMF_subsection}), as well as its mass-stratified sub-samples. The observed stellar population of the embedded Cep~B cluster may be complete somewhere above $1$~M$_{\odot}$ (\S \ref{IMF_subsection}) and thus the analysis of the $>1$~M$_{\odot}$ stellar sub-sample may provide more realistic disk fraction estimates for this sub-region.

Examining first azimuthal dependence, we  find that the NE zone of the region has $\sim 1.5-1.7$ times higher stellar surface density than the SW zone. There is also an indication for possible slight increase in disk fraction from NE to SW zones of the Cep~OB3b/S~155 region, although the effect is not statistically significant (Table \ref{tbl_frac}).

Examining next the radial dependence, a dramatic trend is seen: the disk fraction increases from $40-50$\% in the Cep~OB3b subregion to $50-60$\% in the S~155 subregion to $70-80$\% in Cep~B (Table \ref{tbl_frac}). This trend is elucidated in more detail in Figure \ref{fig_disk_frac_spatial} as a running average disk fraction. Here the running disk fraction for the $>0.5$~M$_{\odot}$ mass stratum (red) is evaluated within a $2\arcmin$ wide sliding window at $1\arcmin$ intervals,  while the $0.5<M<1$~M$_{\odot}$ (black) and $M>1$~M$_{\odot}$ (blue) mass strata are evaluated within a $4\arcmin$ wide sliding window at $2\arcmin$ intervals.  Note these values are not independent because a given star is included in two adjacent values. The effect could either be viewed as a smooth trend in disk fractions, or a step-function at 30\% (outer layer), 60\% (intermediate layer) and $70-80$\% (inner layer) disk fractions in $0.6-0.8$~pc radial bins (grey bars). It is important to note that the outer edge of the second bin (intermediate layer) is coincident with the projected position of the ionizing source of the region, O7V star HD~217086.

If one adopt the simple relation between stellar age and disk fraction derived by \citet[][see their Figure~14]{Hernandez07}, the Cepheus disk fraction gradient corresponds to an age gradient from $3-5$~Myr in the outer layer (part of Cep~OB3b subregion), $2-3$~Myr in the intermediate layer (part of Cep~OB3b/S~155 region), and $\sim 1$~Myr in the inner layer (embedded Cep~B subregion). This inferred age gradient agrees with that estimated from stellar ($vs.$ disk) properties in Appendix~\ref{distance_age_subsection}.

\subsection{Transition Disks  \label{to_subsection}}

As noted earlier in \S~\ref{slope_subsection}, the Cepheus~OB3b population appears deficient in transition disks compared to the $2-3$~Myr old cluster IC~348.  We investigate this further here, including comparison to the low-mass 1~Myr old Coronet cluster \citep{Sicilia-Aguilar08} and the 4~Myr old Tr~37 cluster \citep{Sicilia-Aguilar06}.  Tr~37, like our Cep~OB3b field, harbors an O7-type star.

It is important to recall that different studies of different star forming regions use different criteria for identifying TDs. In the Tr~37 and the Coronet cluster studies, TDs are identified with a photometric excess longward of $\sim 6\mu$m. In the study of IC~348, small excesses around $2$~$\mu$m define ``anemic disks'' systems \citep{Lada06}.  Our classification of Cepheus stars here  (\S \ref{yso_classification_section}) relies on an intermediate criterion of excess longward of $\sim 3.6\mu$m.   Thus, if the fraction of transition disks were intrinsically the same in all clusters, we expect the reported fractions to be largest in IC~348, intermediate in our sample, smallest in Tr~37 and the Coronet.   Furthermore, inter-cluster comparison of TD fractions should also consider similar mass strata, as lower-mass PMS stars may have flatter disks with more frequent inner holes than more massive PMS stars \citep[e.g.][]{Hartmann06, Sicilia-Aguilar08}. From Table~2 of \citet{Lada06}, a TD fraction among disk-bearing stars in IC~348 decreases from $0.45 \pm 0.04$ ($64/142$) for M-type stars to $0.31_{-0.09}^{+0.15}$ ($4/13$) for G-K-type stars, although this decrease is not statistically significant. The Tr~37 study combined stars from late G to M2-type ($\sim 0.4-2$~M$_{\odot}$) and reported a $0.10_{-0.02}^{+0.03}$ ($11-14$ out of $140-150$) TD fraction among disk-bearing stars.  The Coronet cluster study included only M0-M8 objects and reported a $0.50_{-0.12}^{+0.13}$ ($7/14$) TD fraction.

Our classification of X-ray selected PMS stars in the Cep~B/Cep~OB3b obtained 20 TDs out of 139 disk-bearing stars giving a TD fraction of $0.14_{-0.02}^{+0.03}$. Most of these (13 out of 20) are in the unobscured Cep~OB3b subregion and none are present close to the Cep~B hot core. To perform a mass-stratified analysis, we consider here only the most rich star sample, Cep~OB3b. From Table~\ref{tbl2}, we find the Cep~OB3b TD fraction may be mass-dependent with $0.29_{-0.07}^{+0.1}$ (8 out of 28) among $M \la 0.5$~M$_{\odot}$ (M-type) disk-bearing stars and $0.11_{-0.03}^{+0.06}$ (5 out of 47) around more massive stars. 

These results are roughly consistent with the previous studies, once the TD definitions and mass ranges are taken into account.  Our overall $\sim 14\%$ TD fraction is intermediate between the $\sim 40\%$ reported for IC~348 and $\sim 10\%$ reported for Tr~37, as expected for an intrinsically constant TD fraction.  Our finding that M stars have a higher TD fraction than more massive stars agrees with the indication for trend seen in IC~348 and the high fraction found in Coronet M stars. We thus conclude that the fraction of Cep~OB3b TDs agrees with that of other clusters despite the apparent deficit noted earlier in Figure~\ref{fig_slope_histogram}.

\section{IMPLICATIONS FOR TRIGGERED STAR FORMATION \label{discussion_section}}

\subsection{Analogy with Bright Rimmed Clouds \label{BRC_subsection}}

The observed picture of the Cep~B region is in many respects reminiscent, on a larger scale of several parsecs, of those of smaller  bright-rimmed clouds (BRCs) and cometary globules (CGs) found on the edges of giant H~II regions \citep{Sugitani91}. BRCs are isolated clouds surrounded by ionized rims facing the exciting star(s) with their dense cores close to the rims. BRCs are modeled as externally illuminated, photoevaporated and ablated into elongated head-tail morphologies by ultraviolet radiation of OB stars \citep{Reipurth83}. It is likely that pressure from the ionization shock front at the surface propagates through a globule and overcomes the magnetic, turbulent and thermal pressure that supports it against collapse, thereby triggering localized star formation. This astrophysics of radiation-driven implosion (RDI) in molecular globules has been extensively studied \citep[e.g.,][]{Bertoldi89, Lefloch94, Miao09}. RDI models typically consider the triggering of a single star formation episode in a small ($<1$~pc) cloud with tens of M$_{\odot}$ of molecular gas. The characteristic timescale for producing cometary morphologies and inducing gravitational collapse varies with initial conditions from $0.1$ to $\sim 1$~Myr. 

In most cases, molecular, IR, H$\alpha$ surveys of BRCs trace only the most recently formed stars \citep[e.g.,][]{Sugitani95, Ogura02, Thompson04b, Urquhart06}.   In a few cases, $Chandra$ observations have added the diskfree PMS populations \citep{Getman07, Sanchawala07, Getman08a, Wang09}.  Triggered BRCs often show a few embedded mid-IR sources denoting protostars, while H$\alpha$, $JHK$ and X-ray surveys reveal small clusters of disk-bearing PMS stars within and in front of the bright rim.  In a few cases, spatial-age gradients in the stellar population are seen where the youngest stars are embedded and older stars are aligned toward the ionizing sources \citep{Matsuyanagi06, Ogura07, Getman07}.  This directly supports the RDI mechanism and implies that the existing clouds have been actively forming stars for several million years.

The Cepheus region possesses all of the observational features of an RDI triggered star formation region: \begin{enumerate}

\item The presence of exciting star(s) and a molecular cloud surrounded by an ionized rim facing the exciting stars shown in Figures~\ref{fig_map}-\ref{fig_color_image} \citep{Minchin92,Beuther00};

\item The presence of a dense molecular core close to the rim \citep{Yu96,Beuther00};

\item The spatio-temporal gradient of young stars oriented towards the exciting star(s) shown in Figures~\ref{fig_spat_distrib}-\ref{fig_disk_frac_spatial}.

\end{enumerate}
The major difference of the Cepheus region from other BRCs is its larger scale: its extent is $\sim 4$~pc linear size, its molecular mass is  $\sim 200$~M$_{\odot}$, its observed PMS X-ray stellar population is $>60$ embedded and up to $\sim 300$ older unobscured members, as well as estimated intrinsic population of $\sim 1000$ unobscured and possibly a few hundred embedded stars (\S \ref{IMF_subsection}). The large-scale morphology is also different from most other BRCs.  Rather than lying on the edge of a circular H{\sc II} region surrounding a concentrated OB association, it is an extension of a giant molecular cloud protruding into a large evacuated region produced by a partially dispersed OB association.  Another unusual case of a very large BRC with star formation triggered over several million years is the high-latitute cometary globule CG~12 \citep{Getman08a}.

\subsection{Implications for the Radiative Driven Implosion Mechanism \label{RDI_subsection}}

Based on the RDI triggered star formation concept, we can propose some new insights into the star formation process of the Cepheus region and make some simple model-independent estimates for a shock propagation velocity and a star formation efficiency (SFE) of the RDI process. In \S \ref{spatial_domain_subsection} and Appendix~A, we establish the spatio-temporal gradient of Cepheus PMS stars from the center of the cloud towards the HD~217086 O7Vn star.  Assuming a distance around 725~pc, the innermost $0.6-0.8$ parsec adjoining the hot core of the cloud has the youngest ($<1-2$~Myr) stars with the highest seen disk fraction ($70-80\%$).  The intermediate $0.6-1$ parsec layer around S~155 has older ($2-3$~Myr) stars with an intermediate disk fraction ($\sim 60\%$) (Figure \ref{fig_disk_frac_spatial}). The outer layer of mostly diskless stars in the unobscured Cep~OB3b cluster has the lowest disk fraction ($\sim 30\%$)\footnote{If the distance to the cloud is $\sim 850$~pc, the ages of the Cep~B and S~155/Cep~OB3b subregions are older by $\sim 0.5-1$~Myr but the $\sim 1$~Myr difference between them remains (Appendix~A).}

We consider two scenarios for RDI-induced star formation in this region.  First, a relatively slow shock with speed $\la 0.5$~km~s$^{-1}$ may have passed through the molecular cloud, triggering star formation in the outer layers around $2-3$~Myr ago and continuing to trigger star formation close to the Cep~B hot core today. This slow shock speed is consistent with the $0.6$~km~s$^{-1}$ propagation rate inferred from the age gradient of the PMS stars in the globule IC~1396N, located $\sim 11$~pc projected distance of its ionizing O6e type star \citep{Getman07}.  In this scenario, stellar kinematic drift from their birthplaces plays an important role in relocating PMS stars. The O7Vn star HD~217086 star is probably the principal ionization source but other OB stars (Figure \ref{fig_map}), including the B1Vn star HD~217061 currently located within the SW zone of the S~155 subregion, probably also play a role.   If the original cloud were only slightly larger than the size observed today, the RDI shock would have slowly propagated through $\sim 1$~pc of the Cep~B molecular cloud over $\sim 2-3$~Myr years, implying a shock velocity around $\la 0.5$~km~s$^{-1}$.  The stars formed early in this process would have drifted $2-10\arcmin$\/ ($0.4-2$~pc) in all directions, populating much of the $Chandra$ field of view with triggered stars. Such star drifting would require a stellar velocity dispersion around 1~km~s$^{-1}$.  This is expected from the turbulent velocities within a cloud several parsecs in extent \citep{Efremov98}, which can be supplemented by dynamical interactions between protostars during star formation \citep{Bate08, Furesz08}. 

A second scenario involves the passage of a faster shock propagating at $\sim 1$~km~s$^{-1}$ through a cloud that was originally much larger than the Cep~B cloud we see today. In this model, most of the original cloud material has ablated and most of the stars in the $Chandra$ field formed in the cloud during the past $\sim 2-3$~Myr. Present locations of the intermediate layer around S~155 and the inner layer around the molecular core (separated by $0.8-1$~pc, Figure \ref{fig_disk_frac_spatial}) may be directly associated with the passage of the RDI shock $\sim 2-3$ and $\la 1-2$~Myr years ago, respectively. Star drift plays a less important role here. A similar fast shock speed is observed in the triggered star formation in the molecular pillars of the Eagle Nebula (M~16) and other bright rimmed clouds \citep{Fukuda02, Thompson04b}, and is consistent with the theoretical models of \citet{Motoyama07}. 

A critical discriminant between these scenarios is the SFE of the cloud.  Integrating the stellar masses in Figure~\ref{fig_imf}, the total stellar mass of the Cep~OB3b disk-bearing population assuming a standard Galactic field IMF is $\sim 130$~M$_{\odot}$ in our observed field.  With the average disk fraction of $45\%$ (\S \ref{mass_domain_subsection}, Table~\ref{tbl_frac}), we add $\sim 160$~M$_\odot$ of diskfree stars and $\sim 20$~M$_\odot$ from HD~217086 to give a total mass of the Cep~OB3b population around $310$~M$_{\odot}$. A similar estimate for the S~155 population from information in Figure~\ref{fig_imf} and Table~\ref{tbl_frac} gives a total stellar mass around $80$~M$_{\odot}$. Assuming that the observed Cep~B embedded population is complete only somewhere above $1$~M$_{\odot}$ (\S \ref{IMF_subsection}) and follows a standard Galactic field IMF as well, we use star count and disk fraction information for the highest mass bin from Figure~\ref{fig_imf} and Table~\ref{tbl_frac} to obtain an estimated total mass of the Cep~B population of $120-250$~M$_{\odot}$. The mass of the observed molecular gas in the Cep~B molecular portion of the Cepheus molecular cloud is only $\sim 200$~M$_{\odot}$ \citep{Yu96,Beuther00}.

The inferred SFE for the embedded Cep~B population alone is thus around $35\%-55\%$, at the top of typical SFE range of $10-40\%$ measured in other active star forming regions \citep[and references therein]{Elmegreen00}. This is far above the SFE found in smaller molecular globules triggered by UV shocks \citep[e.g.][]{Getman07}. For the total Cepheus stellar population within the $Chandra$ field, the apparent SFE is extremely high around $\ga 70\%$. This is unrealistically high and indicates that most of the Cep~OB3b stars were not formed from the presently seen cloud material but arose from an earlier generation of star formation from gas that is no longer present. This supports the second scenario involving a fast shock passing through a much larger and more massive original cloud.

Two other aspects of RDI triggered star formation can be discussed.  First,  \citet{Sugitani91} provided observational evidence for a non-standard IMF where intermediate-mass stars are preferentially formed over lower mass stars in BRCs. In the case of the embedded triggered PMS population of IC~1396N, \citet{Getman07} also found indications for a nonstandard IMF biased towards higher mass stars, but cautioned about possible observation selection effects. A similar situation arises here in the case of the embedded Cep~B population (\S \ref{IMF_subsection}). While the shape of the Cep~B stellar mass distribution is not definite yet, the stellar mass distribution of likely RDI triggered stellar population of the Cep~OB3b/S~155 region agrees well with a standard Galactic IMF, at least down to $\sim 0.2-0.5$~M$_{\odot}$ (\S \ref{IMF_subsection}). 

Second, most theoretical calculations of RDI triggering involve small globules and a single episode of triggering. The $2-3$~Myr range of stellar ages found in the Cep~B/Cep~OB3b region, and an even wider age spread found in CG~12 \citep{Getman08a}, implies repeated or continuing star formation over millions of years when the RDI mechanism occurs in a larger molecular cloud.

\section{CONCLUSIONS \label{conclusions_section}}

We present a $Spitzer$ IRAC observation of the Cepheus~B molecular cloud, the S~155 HII region on its periphery, and a portion of the nearby Cep~OB3b OB association. The goals of this work are to study disk evolution of the nearly disk-unbiased combined $Chandra$-ACIS and $Spitzer$-IRAC selected samples of PMS stars and to provide new clues of the sequential triggered star formation process in the region.

Out of $\sim 400$ X-ray emitting PMS stars in the region, 354 are classified as diskless or disk-bearing stars based on IR photometry: 161, 36, and 18 (75, 26, and 38) diskless (disk-bearing) stars in the Cep~OB3b, S~155, and Cep~B subregions, respectively. We immediately see that samples selecting only IR-excess stars miss the majority of stars outside the Cep~B molecular core. Among all X-ray emitting disk-bearing systems, only 4 are Class~I protostars; they lie in the younger embedded Cep~B cluster (\S \ref{yso_classification_section}). In addition to the $\sim 400$ X-ray emitting PMS stars in the region, we identify $> 200$ non-$Chandra$ IR-excess low-mass members of the region (\S \ref{non_xray_subsection}). For the lightly-obscured Cep~OB3b and S~155 clusters around the cloud, the combined $Chandra$ and non-$Chandra$ PMS samples are complete down to $0.5$~M$_{\odot}$, while for the embedded Cep~B cluster the observed PMS sample may be complete only above $1$~M$_{\odot}$ (\S \ref{IMF_subsection}).

The spatial distribution of the disks around stars of the combined $Chandra-Spitzer$ PMS samples reveals a dramatic spatio-temporal gradient:  younger stars are clustered around the hot core of the molecular cloud, while older stars are dispersed in the direction of the primary ionizing O7-type star HD~217086 (\S \ref{spatial_domain_subsection}). Disk fractions fall from 80\% to 30\% across the S~155 HII region. This finding constitutes strong endorsement for the basic model of triggered star formation in the region by passage of shocks driven by the OB stellar ionization and/or winds into the Cep~B molecular cloud. Our estimate of the shock velocity of $\sim 1$~km~s$^{-1}$ agrees with empirical estimates obtained for a number of other shocked bright-rimmed clouds and with theoretical models. The IMF of the Cep~OB3b/S~155 triggered stellar population exhibits a standard Galactic field shape; we do not confirm our earlier suggestion of a bottom-heavy IMF. The unusually large apparent value of star formation efficiency of $\ga 70\%$ based on the present day molecular cloud indicates that most of the Cep~B molecular material has likely already been evaporated or transformed to stars. Contrary to the current RDI simulations of typical small globules which model a single triggering event over $\la 1$~Myr time period, our observational findings support a picture of multiple episodes of triggering over millions of years (\S \ref{discussion_section}).

Using the nearly disk-unbiased combined $Chandra$-ACIS and $Spitzer$-IRAC selected stellar sample, we also obtain several other results. First, the disk fraction as a function of stellar mass in the Cep~OB3b cluster is similar to that of the IC~348 PMS stars with ages similar to Cep~OB3b (\S \ref{mass_domain_subsection}). Second, the fraction of Cep~OB3b transition disks agrees with that of other clusters, such as Tr~37, Coronet, and IC~348 (\S \ref{to_subsection}). Third, in agreement with the disk fraction analysis of \S \ref{spatial_domain_subsection}, we obtain photometric ages of $<1-2$ and $2-3$~Myr for the embedded Cep~B and lightly-obscured Cep~OB3b clusters, respectively using the optical color-magnitude diagram (Appendix~\ref{distance_age_subsection}) and assuming a distance of $725$~pc. Fourth, the effect of reduced X-ray emission in disk-bearing versus diskless stars which was originally established in studies of ONC and Taurus regions is confirmed in the Cepheus systems (Appendix~\ref{lx_disks_section}).

\appendix
\section{APPENDIX A: DISTANCE AND AGE OF THE CEP~B/CEP~OB3 COMPLEX \label{distance_age_subsection}}

The distance to the Cep~B/Cep~OB3 complex is somewhat uncertain. From the optical photometry of the early-type members of the Cep~OB3 association, \citet{Blaauw59} and \citet{Crawford70} obtained a distance estimate of $725$~pc. Later \citet{Moreno-Corral93} combined near-IR photometry with the earlier optical photometry and spectroscopy data of bright members to derived a distance estimate of $850$~pc.  A recent VLBI parallax measurement using the methanol maser in the Cep~A molecular core gives a distance of $\sim 700$~pc \citep{Moscadelli08}.   A distance of 870~pc has been inferred for the nearby Cep~OB2 association \citep{Contreras02,Sicilia-Aguilar05}.    Our X-ray study \citep{Getman06} adopted the $725$~pc estimate while the optical studies of the Cep~OB3b by \citet{Pozzo03} and \citet{Mayne07} adopted a distance of $850$~pc.

The chosen distance influences age estimates of the lower-mass stars as PMS evolutionary isochrones on the Hertzsprung-Russell diagram (HRD) depend on absolute luminosities.  Early photometric age estimates of the OB population indicated a relatively old cluster around $\sim 4-5.5$~Myr \citep{Blaauw64, Jordi96}. For the X-ray PMS stars in the $Chandra$ field, we adopted an age of $\sim 1$~Myr \citep{Getman06} based on the HRD of five PMS X-ray sources \citep{Pozzo03}. Recently, \citet{Mayne07} obtained $V$- and $I_c$-band photometry data for an area including most of the $Chandra$ ACIS-I field and measured a $\sim 3$~Myr photometric age for the low-mass Cep~OB3b population.  It is unclear from any of these earlier datasets whether the age is uniform across the region.

We seek here to estimate stellar ages from the $V$- and $I_c$-band color-magnitude diagram for the $Chandra$ X-ray PMS stars using the photometry of \citet{Mayne07} listed in Table \ref{tbl1}.  The $2MASS$ near-IR photometry was first used with the combination of the synthesized \citet{Siess00} (for $1.4 \leqslant M \leqslant 7.0$~M$_{\odot}$) and \citet{Baraffe98} (for $0.02 \leqslant M \leqslant 1.4$~M$_{\odot}$) PMS evolutionary isochrones to obtain individual extinction estimates for $Chandra$ sources. We obtained different sets of source extinctions ($A_{V,NIR}$) using the $J$ versus $J-H$ color-magnitude diagram \citep[similar to the one shown in Figure 5 of][]{Getman06} with two trial assumed distances (725 and 850~pc) and three trial assumed ages (1, 2, and 3~Myr).  The six sets of source extinctions were applied to correct the observed optical magnitudes of $Chandra$ sources and to place them on the $V$ versus $V-I_c$ diagram.  Based on the values of $E(B-V) = 0.91$~mag and $E(V-I)=1.18$~mag at $A_V = 2.8$~mag adopted by \citet{Pozzo03} and the reddening laws of \citet{Winkler97}, we used the reddening relationships $A_{I_c}/A_V = 0.58$, $A_J/A_V = 0.27$, and $A_H/A_V =0.16$.  The Cep~B optical stars have an average extinction $<A_{V,NIR}> \sim 7$~mag while S~155 stars have $<A_{V,NIR}> \sim 3$~mag and Cep~OB3b stars have $<A_{V,NIR}> \sim 2$~mag.  Within each of the areas, the extinctions of disk-bearing stars are systematically higher than those of diskless stars.  

As the inferred visual source extinctions differ by $A_V< 0.5-1$~mag for the six trial combinations of distance and age, and the reddening vector on the optical color-magnitude diagram is almost parallel to the PMS evolutionary tracks, the X-ray sources occupy very similar loci in the $V-I_c$ color-magnitude diagram for all six cases.  The X-ray sources are centered mostly around the $\sim 2-3$~Myr isochrons assuming the distance of 725~pc, or the $\sim 2$~Myr isochrone assuming the distance of 850~pc.  Figure~\ref{fig_cmd} shows the color-magnitude diagram for 725~pc. The plotted symbols stratify the 286 X-ray sources with optical photometry according to location in the star forming region and the presence of disks: 62 disk-bearing (143 diskless), 22 disk-bearing (31 diskless), and 17 disk-bearing (11 diskless) for the Cep~OB3b, S~155, and Cep~B subregions, respectively.

In Figure~\ref{fig_cmd}, about thirty X-ray sources have $V-(V-I_c)$ locations inconsistent with any PMS model tracks. These outliers are not foreground or background stellar contaminants, as more than half of them have optically thick disks and very few field stars are predicted to be captured in the $Chandra$ image.  Most of these outliers have unusually high near-IR extinctions $A_{V,NIR}$ among the X-ray young stars seen in the optical bands.  For the majority of the outliers, the $A_{V,NIR}$ values are comparable (within $30\%$) to the absorption $A_{V,Xray}$ inferred from the X-ray median energy, $MedE$.  $MedE$ is first converted to equivalent hydrogen column density assuming solar abundances using the $MedE-N_H$ relation found for the COUP sample \citep{Feigelson05}, and $N_H$ is converted to $A_{V,Xray}$ assuming a gas-to-dust ratio of $\sim 2 \times 10^{21}$~cm$^{-2}/$mag \citep{Ryter96}.  This may point to discrepant optical magnitudes and/or unusual conditions in these PMS systems; for example, some may posses disk at high inclination where the optical light is enhanced by scattering above the disk.  Alternatively, these outliers may have ages overestimated by incorrect consideration of birth line and accretion effects  \citep[e.g.][]{Hartmann03}.

From a comparison of soft X-ray absorptions and fractions of $K_s$-band excess sources, \citet{Getman06} presented evidence that the embedded Cep~B cluster is younger than the Cep~OB3b population within the $Chandra$ ACIS-I field. Figure~\ref{fig_cmd} supports this age difference.  Ignoring outliers mentioned above, about half of the Cep~B sources (blue symbols) are located closer to the 1~Myr track compared to only about one-sixth of the Cep~OB3b (red symbols). However, the Cep~B sample is small because most of the X-ray selected members are embedded and do not have optical counterparts. The result supports the youthfulness of the Cep~B population seen in the disk fraction discussed in \S~\ref{spatial_domain_subsection}.

We thus cannot arrive at a clear conclusion on the distance to the star forming complex.  However we do establish a dependence of the inferred average age on the assumed distance.  In this paper, we adopt (not derive) a distance to the Cep~B/Cep~OB3b complex of 725~pc, supported by the recent VLBI parallax measurement of Cep~A.  We therefore adopt an age of $2-3$~Myr for the Cep~OB3b stars and $<1-2$~Myr for the embedded Cep~B stars.  If the distance is closer to 850~pc, the inferred ages would be decreased by $\sim 0.5-1$~Myr.

\section{APPENDIX B: X-RAY ACTIVITY, STELLAR MASS AND DISKS \label{lx_disks_section}}

Originally found by \citet{Feigelson93} from ROSAT data, it is now empirically well-established that PMS X-ray luminosities are strongly correlated with stellar mass, volume and surface area.  The $L_x \propto M^{1.7}$ relationship extends over 3 orders of magnitude range in X-ray luminosity.  The clearest relationships are seen in the {\it Chandra Orion Ultradeep Project} (COUP) observation of the Orion Nebula Cluster \citep{Getman05, Preibisch05b} and in the {\it XMM-Newton Extended Survey of Taurus} \citep[XEST,][]{Gudel07, Telleschi07}.  The astrophysical cause of this relationship is poorly understood, but probably is due to saturation of the magnetic dynamo in the fully convective stellar interior or on the surface of PMS stars.  

\begin{figure}
\centering
\includegraphics[angle=0.,width=5.5in]{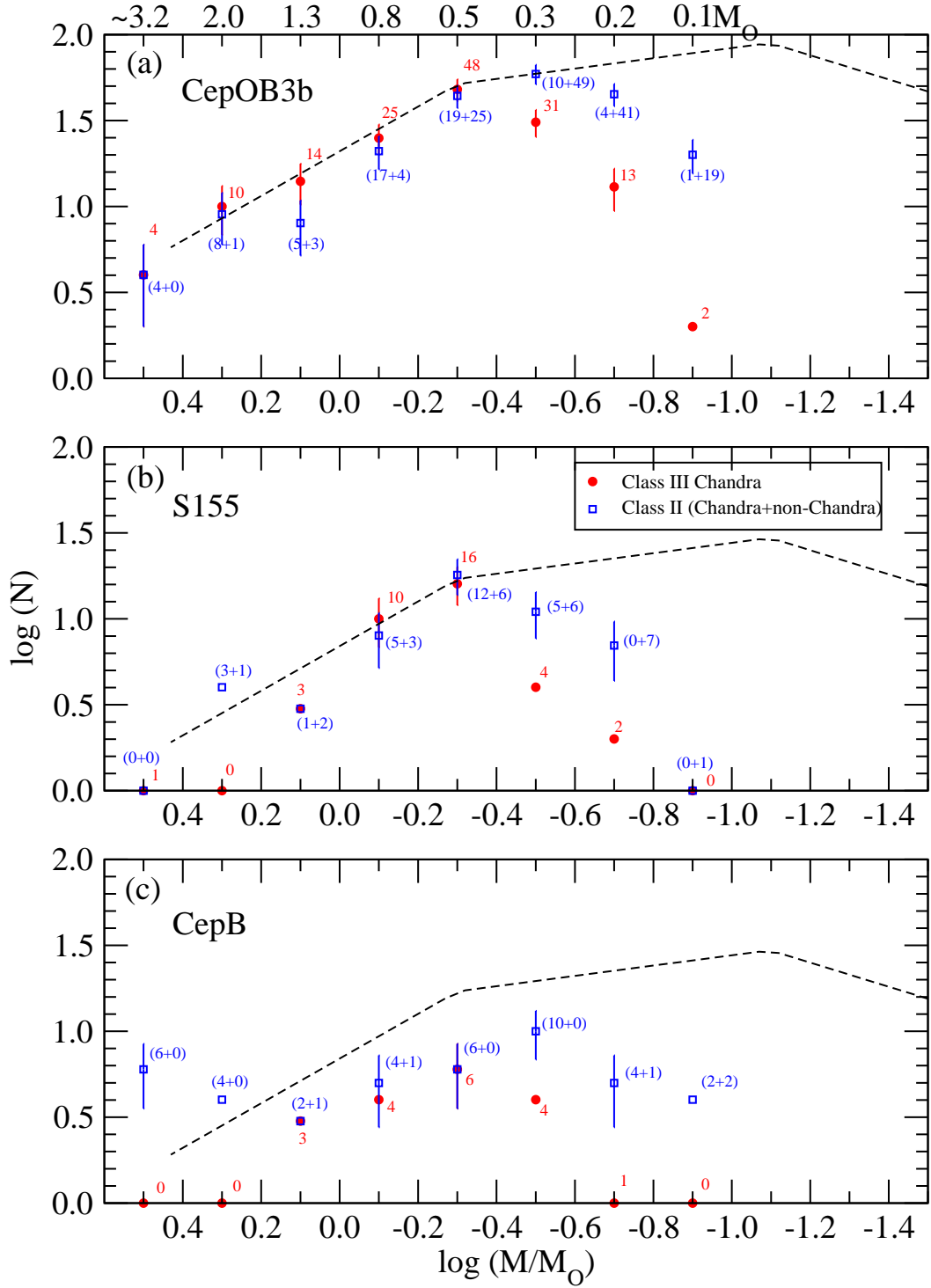}
\caption{Mass distribution of Class~III $Chandra$ (red) and combined Class~II $Chandra$ and non-$Chandra$ (blue) PMS stars in the  (a) Cep~OB3b, (b) S~155, and (c) CepB sub-regions. The numbers of stars are estimated within mass bins of 0.2~dex width and $1\sigma$ equivalent Poisson errors are shown. Annotation labels indicate the corresponding to each bin number of detected $Chandra$ Class~III stars (red), as well as $Chandra$ $+$ non-$Chandra$ Class~II stars (blue). The dashed lines represent the scaled versions of the Galactic field IMF \citep{Kroupa02}. \label{fig_imf}}
\end{figure}

\begin{figure}
\centering
\includegraphics[angle=0.,width=6.5in]{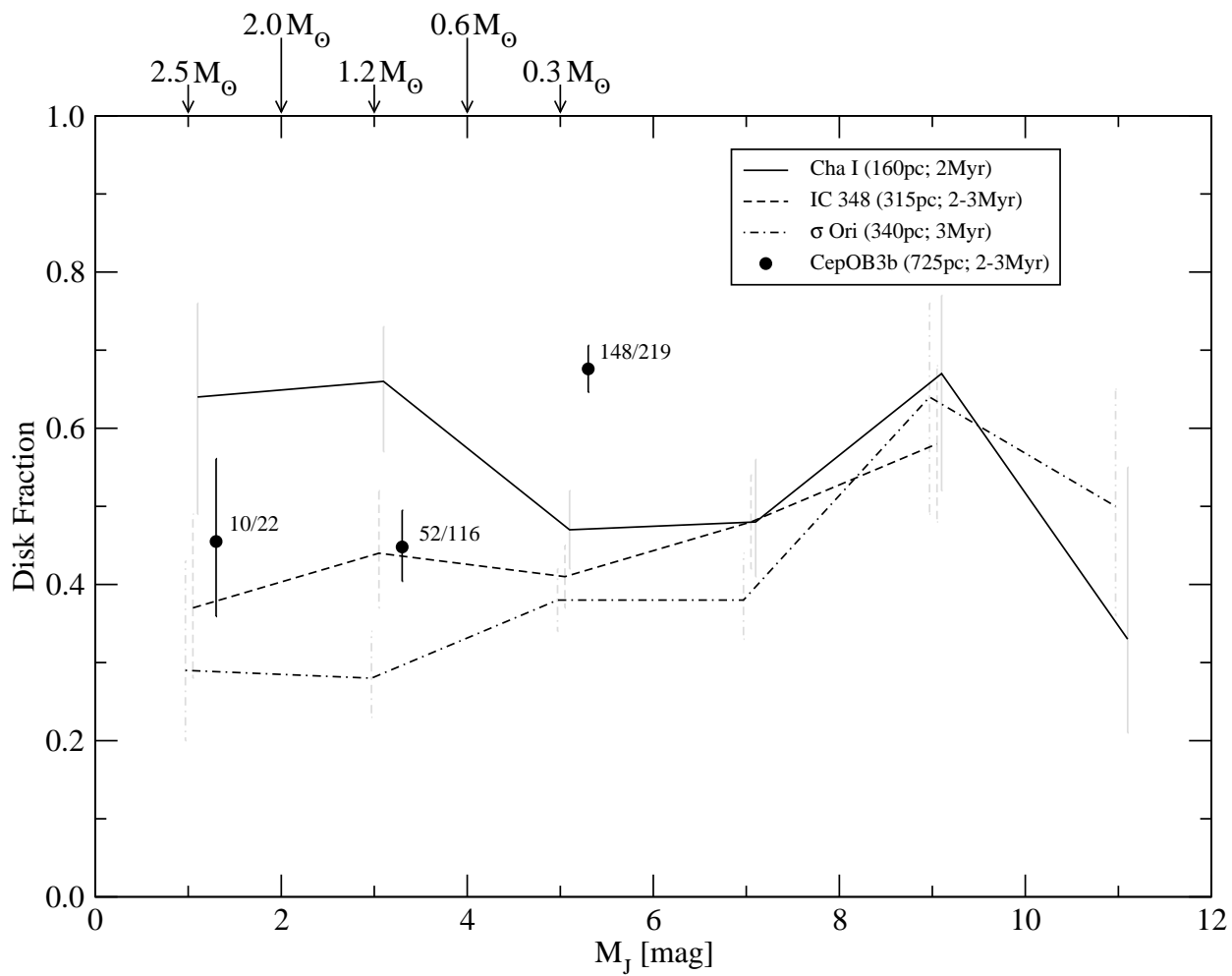}
\caption{Comparison of disk fraction as a function of stellar absolute $J$-band magnitude for four $2-3$~Myr stellar clusters:  Cep~OB3b (circles with 1$\sigma$ error bars); Cha~I (solid line); IC~348 (dashed line); and $\sigma$~Ori (dashed-dotted line).  The Cep~OB3b stellar sample is the combined Class~II/III $Chandra$ and Class~II non-$Chandra$ PMS stars from this work. Data for the three other clusters are from \citet{Luhman08b}. The top axis shows an approximate scaling to stellar mass. \label{fig_mj_diskfraction}}
\end{figure}

\begin{figure}
\centering
\includegraphics[angle=0.,width=7.0in]{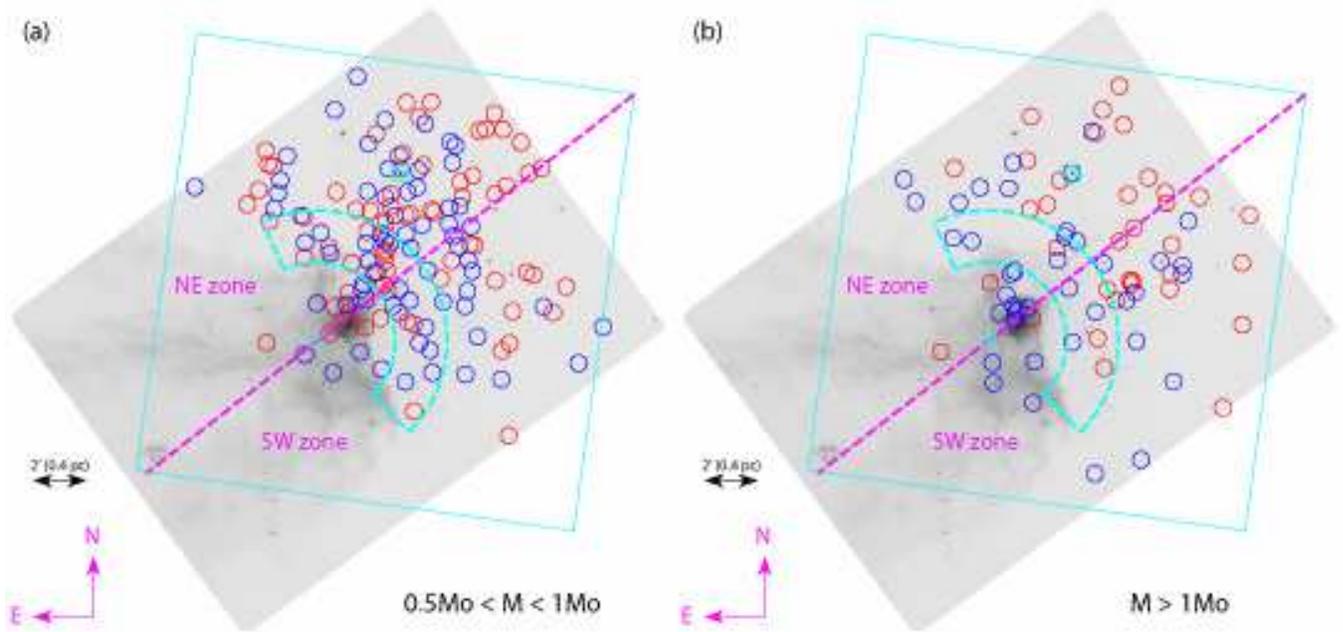}
\caption{Spatial distribution of Cepheus $Chandra$ plus non-$Chandra$ mass-complete selected PMS samples superposed on a grey-scale 5.8~$\mu$m band $Spitzer$-IRAC image:  diskfree stars (red), disk-bearing stars (blue).  Panel  (a) shows $0.5 \la M \la 1.0$~M$_{\odot}$ stars, and panel (b) shows $M \ga 1.0$~M$_{\odot}$ stars. The O7Vn star HD~2170786 is marked by the cyan square-circle, and the molecular hot core by the cyan cross. The cyan dashed locus delineates the S~155 subregion. The $Chandra$ field of view is outlined by the large cyan square, and the NE and SW azimuthal zones are demarcated by the magenta line. \label{fig_spat_distrib}}
\end{figure}

\begin{figure}
\centering
\includegraphics[angle=0.,width=7.0in]{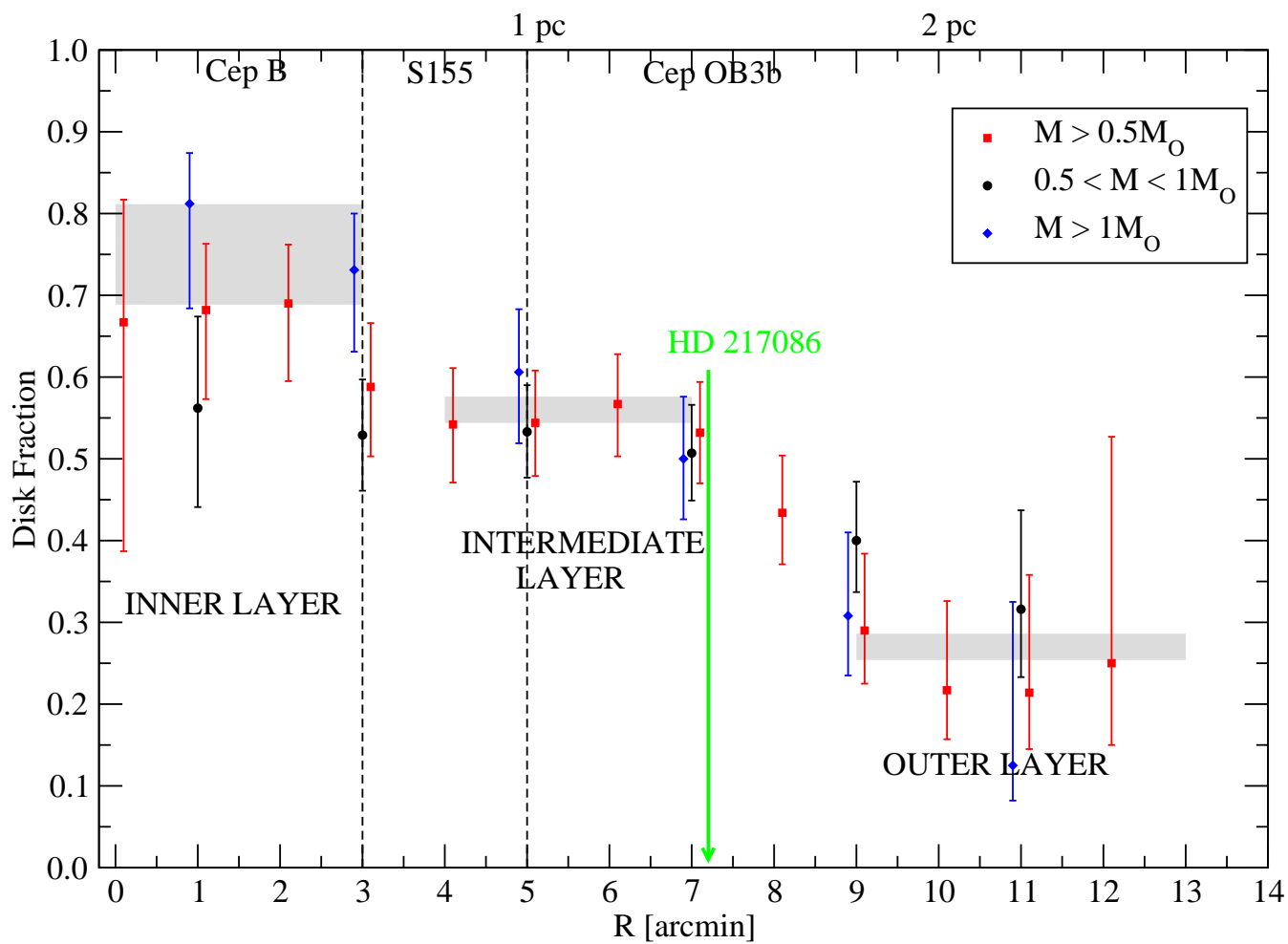}
\caption{Running average of the disk fraction as a function of angular distance from Cep~B hot core for the combined $Chandra$ plus non-$Chandra$ sample of PMS stars. The three subregions are separated by dashed lines and the green vertical arrow indicates the projected location of the primary ionizing star HD~217086. Mass-stratified sub-samples are color coded: $M > 0.5$~M$_{\odot}$ (red), $0.5 < M < 1$~M$_{\odot}$ (black), and $M > 1$~M$_{\odot}$ (blue). An interpretation based on three constant disk fraction layers is indicated by horizontal grey bars with their disk fraction levels consistent with those from Table~\ref{tbl_frac}. \label{fig_disk_frac_spatial}}
\end{figure}

\begin{figure}
\centering
\includegraphics[angle=0.,width=6.5in]{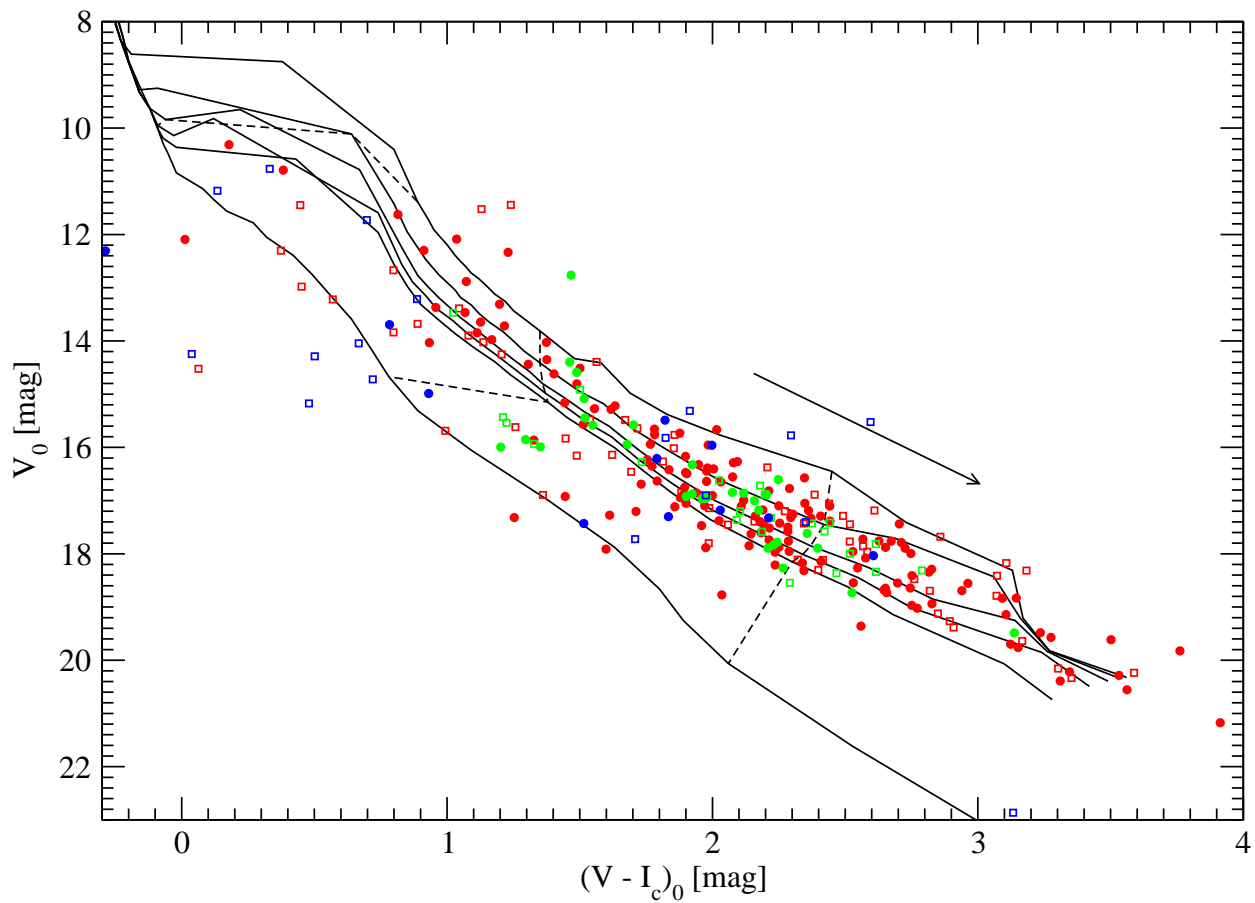}
\caption{$VI_c$ color-magnitude diagram showing the Cep~B/Cep~OB3b X-ray selected PMS stars individually corrected for extinction. Disk-bearing stars are plotted as open squares, and diskfree stars as filled circles. Colors indicate subregion location: Cep~OB3b (red), S~155 (green), and Cep~B (blue).  Solid lines show PMS isochrones for ages 1, 2, 3, 4, 5~Myr and the Zero-Age Main Sequence from evolutionary tracks of \citet{Siess00} assuming $d=725$~pc.  Dashed lines show mass tracks 0.3, 1.0, and 3.0~M$_{\odot}$.  The reddening vector shows $A_V \sim 2$~mag. \label{fig_cmd}}
\end{figure}

\begin{figure}
\centering
\includegraphics[angle=0.,width=7.0in]{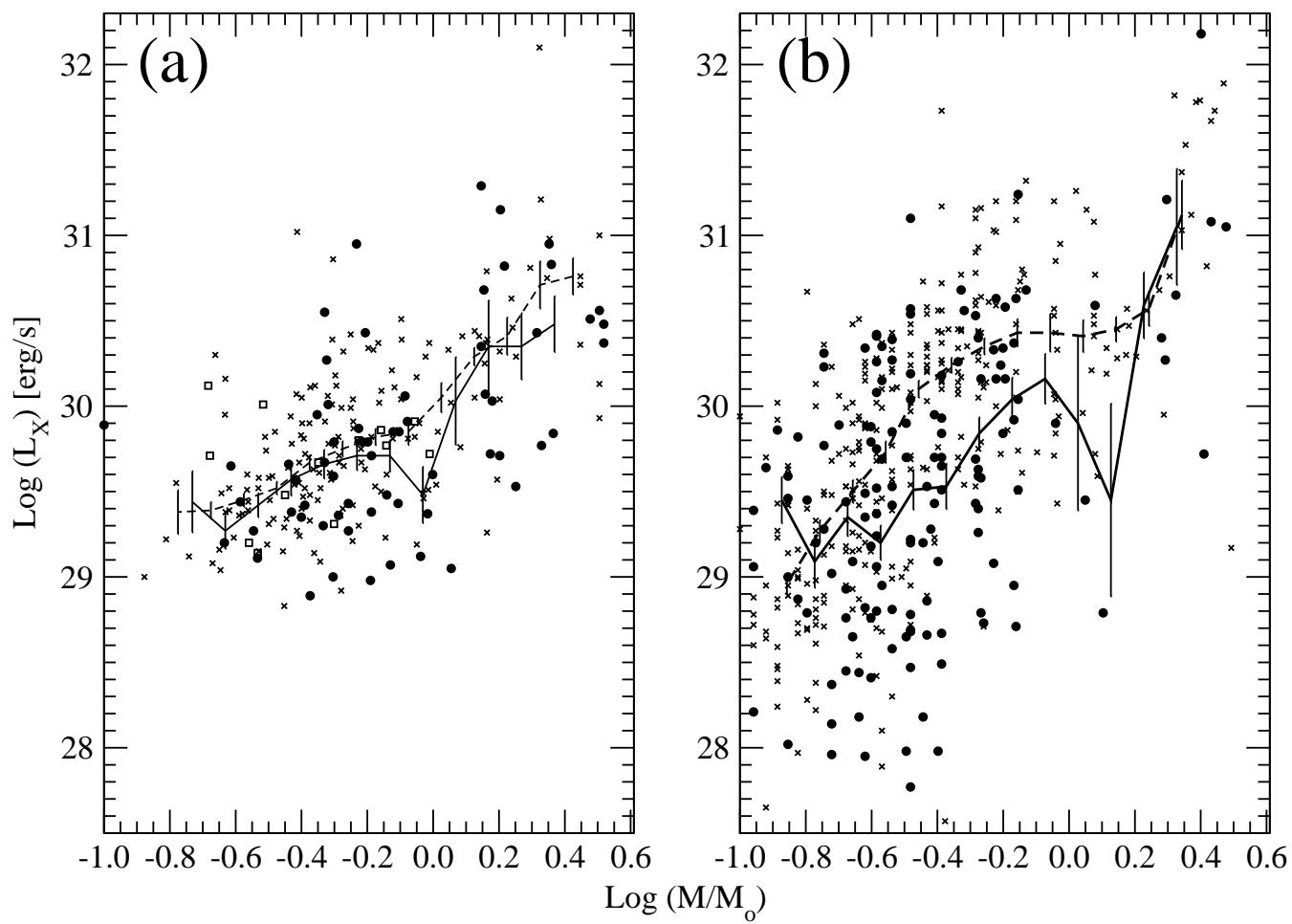}
\caption{X-ray luminosity as a function of mass for lightly-obscured PMS stars in (a) Cep~OB3b and (b) the Orion Nebula Cluster from \citet{Getman05}. In this figure, $\times$ symbols are diskless or non-accreting stars, {\large\bf $\bullet$} are optically thick disk or accreting stars, and {\large\bf $\Box$} are transition disk objects.  Running medians for $\log L_x$ are shown as dashed (solid) lines for diskless or non-accreting (optically thick disk or accreting) stars. \label{fig_lx_vs_mass}}
\end{figure}

Figure \ref{fig_lx_vs_mass} shows the $L_x -M$ relationship for 222 unobscured Cep~OB3b PMS stars compared to 457 lightly-absorbed ($A_V < 5$~mag) Orion COUP stars from \citet{Getman05}.  Here, $L_x$ represents the quantity $L_{t,c}$, the X-ray luminosity computed in the $0.5-8$~keV band and corrected for obscuration.  A similar, roughly linear, relationship is seen in the Cep~OB3b sample, but somewhat flatter than the relationship present in the COUP sample.  This can be attributed to truncation effects: the Cep~OB3b observation is complete to $\log L_x \simeq 29.5$ erg~s$^{-1}$ \citep{Getman06}, while the COUP observation is complete to $\log L_x \simeq 28.0$ erg~s$^{-1}$. Examination of Figure~\ref{fig_lx_vs_mass} shows that the Cep~OB3b X-ray sample alone is nearly complete down to $0.5$~M$_{\odot}$ and $\sim 1$~M$_{\odot}$ for the Class~III and Class~II stars, respectively (see Figure \ref{fig_imf}).  Much of the scatter in both relations can be attributed to errors in COUP and Cep~OB3 stellar masses; less scatter is seen in the XEST $L_x-M$ plot where masses are carefully evaluated from accurate spectroscopy. X-ray members of the embedded Cep~B cluster show a similar $L_x - M$ relationship with the completeness limit above $1$~M$_{\odot}$.

A second relationship between X-ray emission and accretion was originally found from $Chandra$ HRC (High Resolution Camera) data \citep{Flaccomio03} and has been confirmed in the COUP and XEST samples. This is thought to arise from a suppression of time-integrated X-ray emission in accreting versus non-accreting PMS systems. This is a more subtle effect than the X-ray dependence on mass with an amplitude of only a factor of $\sim 2$ within the $10^3$ range in PMS X-ray luminosities \citep{Preibisch05b, Telleschi07}.   The astrophysical cause of the mild suppression of X-ray emission in accreting PMS systems is also uncertain.  Preibisch et al. suggest that X-ray emission cannot arise in magnetic field lines that are mass loaded with disk material. \citet{Jardine06} argue that the outer magnetosphere of accretors is stripped by interaction with the disk. \citet{Gregory07} propose that soft X-ray emission is attenuated by dense material in accretion columns. \citet{Getman08c} find indications of shorter X-ray flare durations in Class~II systems that may be due to distortion and destabilization of magnetic loop structures in accreting systems.

This effect is shown in Figure~\ref{fig_lx_vs_mass} with running medians of accreting and non-accreting stars for the lightly-obscured Cep~OB3b and COUP samples.  The running medians are calculated within a sliding window with width 0.3~dex in $\log (M/M_{\odot})$. The errors on medians are median absolute deviations \citep[see Appendix~B of][]{Getman08b}.  The sample selections are not identical due to different X-ray sensitivity limits and due to different limited spectroscopic/photometric measurements in Orion and Cep~OB3b regions.  We show with the {\large\bf $\bullet$} symbol 62 Cep~OB3b stars classified as disk-bearing  `DSK' systems in \S~\ref{yso_classification_section}, and 142 COUP stars classified as active accretors having Ca~II 8542~$\AA$ line in emission with equivalent width of $EW {\rm (Ca~II)} < -1\AA$. The $\times$ symbols indicate 160 Cep~OB3b stars classified as diskless `NoD' and 315 weakly accreting or non-accreting COUP stars with absorption equivalent width of $EW \rm{(Ca~II)} > 1\AA$. Thirteen Cep~OB3b transition disk stars are marked by squares.  

The factor of $\sim 2$ offset between the accreting and non-accreting stars in the COUP samples is clearly seen to be superposed on the dominant $L_x-M$ relationship, but the effect is only marginally present in the Cep~OB3b sample.  The weakness of the offset here can be attributed to two effects.  First, the incompleteness at lower-$L_x$ values discussed above has a greater impact on the weaker-$L_x$ accreting stars than the stronger-$L_x$ non-accreting stars. Second, due to the absence of a spectroscopic survey of Cep~OB3b stars, we classify stars here using an infrared photometric disk indicator rather than a direct accretion indicator.  It has been established within the COUP sample that the presence of a disk itself does not suppress X-ray emission, but active accretion must be present \citep{Preibisch05b}.  These two effects plausibly explain the less pronounced observed effect of the Cep~OB3b Class~II X-ray emission suppression compared to that of the COUP stars.

\acknowledgements We thank the anonymous referee for very useful comments that improved this work. We thank Robert Gutermuth (CfA) for development of Spitzer data analysis tools, and Patrick Broos and Leisa Townsley (Penn State) for their development of $Chandra$ tools.  We benefited from scientific discussions with Jeroen Bouwman, Thomas Henning and colleagues at MPIA.  This work is supported by the Chandra ACIS Team (G. Garmire, PI) through the SAO grant SV4-74018 and by Guest Observer JPL grant \# 1287737 (J. Wang, PI). K. L. was supported by grant AST-0544588 from the National Science Foundation. This work is based on observations made with the {\it Spitzer Space Telescope}, which is operated by the Jet Propulsion Laboratory, California Institute of Technology under a contract with NASA. We also use data products of the Two Micron All Sky Survey, which is a joint project of the University of Massachusetts and the Infrared Processing and Analysis Center/California Institute of Technology, funded by NASA and NSF.



\clearpage
\clearpage

\begin{deluxetable}{ccccccccccccccc}
\centering \rotate \tabletypesize{\tiny} \tablewidth{0pt}
\tablecolumns{15}
\tablecaption{Optical and IR Photometry of X-ray Sources \label{tbl1}}
\tablehead{

\colhead{No.} & \colhead{Off$_1$} & \colhead{$V$} &
\colhead{$V-I_c$} & \colhead{Off$_2$} & \colhead{$J$} &
\colhead{$H$} & \colhead{$K_s$} &
\colhead{F$_1$} & \colhead{Off$_3$} &
\colhead{[3.6]} & \colhead{[4.5]} &
\colhead{[5.8]} & \colhead{[8.0]} & \colhead{F$_2$}\\

&($\arcsec$)&(mag)&(mag)&($\arcsec$)&(mag)&(mag)&(mag)&&($\arcsec$)&(mag)&(mag)&(mag)&(mag)&\\

(1)&(2)&(3)&(4)&(5)&(6)&(7)&(8)&(9)&(10)&(11)&(12)&(13)&(14)&(15)}

\startdata
280 &   0.1 & $ 21.57 \pm   0.01$ & $  3.86 \pm   0.02$ &   0.0 & $ 15.38 \pm   0.06$ & $ 14.39 \pm   0.08$ & $ 13.74 \pm   0.05$ & AAA000 &   0.1 & $ 12.99 \pm   0.02$ & $ 12.58 \pm   0.02$ & $ 12.36 \pm   0.04$ & $ 11.72 \pm   0.04$ & 4444\\
281 &   0.4 & $ 19.75 \pm   0.01$ & $  3.42 \pm   0.01$ &   0.4 & $ 14.30 \pm   0.03$ & $ 13.31 \pm   0.03$ & $ 12.97 \pm   0.03$ & AAA000 &   0.5 & $ 12.79 \pm   0.02$ & $ 12.70 \pm   0.02$ & $ 12.65 \pm   0.04$ & $ 12.72 \pm   0.06$ & 4444\\
282 &   0.3 & $ 17.30 \pm   0.01$ & $  2.44 \pm   0.01$ &   0.3 & $ 13.06 \pm   0.02$ & $ 12.10 \pm   0.02$ & $ 11.64 \pm   0.02$ & AAA000 &   0.1 & $ 10.76 \pm   0.02$ & $ 10.44 \pm   0.02$ & $ 10.04 \pm   0.04$ & $  9.36 \pm   0.05$ & 4444\\
283 &   0.1 & $ 18.72 \pm   0.01$ & $  3.11 \pm   0.01$ &   0.1 & $ 13.63 \pm   0.03$ & $ 12.63 \pm   0.03$ & $ 12.29 \pm   0.03$ & AAAs00 &   0.1 & $ 12.04 \pm   0.02$ & $ 11.94 \pm   0.02$ & $ 11.97 \pm   0.03$ & $ 12.11 \pm   0.09$ & 4444\\
284 &   0.3 & $ 20.24 \pm   0.01$ & $  3.52 \pm   0.01$ &   0.3 & $ 14.46 \pm   0.03$ & $ 13.50 \pm   0.04$ & $ 13.13 \pm   0.03$ & AAAc00 &   0.2 & $ 12.86 \pm   0.04$ & $ 12.81 \pm   0.04$ & $ 12.72 \pm   0.05$ & $ 12.74 \pm   0.08$ & 2222\\
285 & \nodata & \nodata & \nodata &   0.3 & $ 14.48 \pm   0.04$ & $ 13.47 \pm   0.04$ & $ 13.06 \pm   0.04$ & AAAccc &   0.2 & $ 12.16 \pm   0.02$ & $ 11.79 \pm   0.03$ & $ 11.40 \pm   0.04$ & $ 10.71 \pm   0.05$ & 3333\\
286 &   0.2 & $ 19.76 \pm   0.01$ & $  3.43 \pm   0.01$ &   0.2 & $ 14.11 \pm   0.03$ & $ 13.11 \pm   0.02$ & $ 12.66 \pm   0.02$ & AAA000 &   0.2 & $ 12.15 \pm   0.02$ & $ 11.92 \pm   0.02$ & $ 11.71 \pm   0.03$ & $ 10.99 \pm   0.03$ & 4444\\
287 &   0.4 & $ 17.76 \pm   0.01$ & $  2.40 \pm   0.01$ &   0.5 & $ 13.38 \pm $ \nodata & $ 12.54 \pm $ \nodata & $ 12.26 \pm   0.03$ & UUA00c &   0.6 & $ 12.02 \pm   0.02$ & $ 11.96 \pm   0.03$ & $ 12.18 \pm   0.18$ & \nodata & 3349\\
288 &   0.0 & $ 19.09 \pm   0.01$ & $  3.05 \pm   0.01$ &   0.1 & $ 13.69 \pm   0.02$ & $ 12.50 \pm   0.02$ & $ 12.06 \pm   0.02$ & AAA000 &   0.2 & $ 11.68 \pm   0.02$ & $ 11.62 \pm   0.02$ & $ 11.78 \pm   0.09$ & \nodata & 4449\\
289 &   1.3 & $ 12.55 \pm   0.01$ & $  1.05 \pm   0.02$ &   1.6 & $ 10.79 \pm   0.02$ & $ 10.49 \pm   0.02$ & $ 10.40 \pm   0.02$ & AAA000 &   1.6 & $ 10.33 \pm   0.02$ & $ 10.32 \pm   0.02$ & $ 10.34 \pm   0.03$ & $ 10.37 \pm   0.03$ & 4444\\
290 &   0.2 & $ 19.33 \pm   0.01$ & $  3.03 \pm   0.01$ &   0.3 & $ 14.63 \pm   0.06$ & $ 14.03 \pm   0.05$ & $ 13.80 \pm   0.06$ & AAA000 &   0.3 & $ 13.46 \pm   0.04$ & $ 13.30 \pm   0.05$ & \nodata & \nodata & 2299\\
291 &   0.5 & $ 21.32 \pm   0.01$ & $  3.72 \pm   0.02$ &   0.5 & $ 15.39 \pm   0.07$ & $ 14.49 \pm   0.06$ & $ 14.13 \pm   0.07$ & AAA000 &   0.3 & $ 13.81 \pm   0.02$ & $ 13.69 \pm   0.02$ & $ 13.73 \pm   0.06$ & $ 13.17 \pm   0.09$ & 4444\\
292 &   0.4 & $ 19.26 \pm   0.01$ & $  2.99 \pm   0.01$ &   0.4 & $ 14.06 \pm   0.04$ & $ 12.93 \pm   0.04$ & $ 12.53 \pm   0.04$ & AAA000 &   0.3 & $ 12.16 \pm   0.02$ & $ 12.13 \pm   0.02$ & $ 11.62 \pm   0.05$ & \nodata & 4449\\
293 &   0.4 & $ 18.02 \pm   0.01$ & $  2.66 \pm   0.01$ &   0.4 & $ 13.48 \pm   0.04$ & $ 12.51 \pm   0.03$ & $ 12.24 \pm   0.03$ & AAA0s0 &   0.3 & $ 11.98 \pm   0.02$ & $ 11.94 \pm   0.02$ & $ 11.88 \pm   0.03$ & $ 11.98 \pm   0.03$ & 4444\\
294 &   0.3 & $ 15.88 \pm   0.00$ & $  2.01 \pm   0.01$ &   0.5 & $ 11.71 \pm   0.02$ & $ 10.93 \pm   0.02$ & $ 10.62 \pm   0.02$ & AAA000 &   0.3 & $ 10.35 \pm   0.02$ & $ 10.34 \pm   0.02$ & $ 10.34 \pm   0.03$ & $ 10.27 \pm   0.03$ & 4444\\
295 &   0.1 & $ 20.11 \pm   0.01$ & $  3.49 \pm   0.01$ &   0.2 & $ 14.21 \pm   0.04$ & $ 13.10 \pm   0.03$ & $ 12.62 \pm   0.03$ & AAA000 &   0.5 & $ 12.29 \pm   0.07$ & $ 12.07 \pm   0.07$ & $ 11.86 \pm   0.07$ & \nodata & 2229\\
296 & \nodata & \nodata & \nodata & \nodata & \nodata & \nodata & \nodata & \nodata & \nodata & \nodata & \nodata & \nodata & \nodata & 9999\\
\enddata

\tablecomments{This table is available in its entirety in the electronic edition of the journal. A portion is shown here for guidance regarding its form and content. Column 1: X-ray source number. For X-ray source positions see Table~1 of \citet{Getman06}. Columns 2-4: Optical-X-ray positional offset, and optical $V$, $V-I_c$ magnitudes. Optical data are from \citet{Mayne07}. Columns 5-9: 2MASS-X-ray positional offset, 2MASS $JHK_s$ magnitudes, and 2MASS photometry quality and confusion-contamination flags. For 2MASS source names, see Table~2 of \citet{Getman06}. Column 10: IRAC-X-ray positional offset. For the most of the sources the reported offset is for an IRAC source from the $3.6$~$\mu$m-band image; when not available, the $4.5$~$\mu$m-band position is used. Columns 11-14: IRAC magnitudes derived in this work. Column 15: Four digit flag (one for each IRAC band) giving photometric apertures and level of source contamination from nearby sources and nebular IR emission: 2~pixel aperture with likely high level of contamination, 3~pixel aperture with likely moderate contamination, 4~pixel aperture with likely low level of source contamination. A `9' indicates the inability to derive photometry for one of several reasons: out of IRAC channel field of view (this is further clarified by the flag F$_3$ from Table~2), weak or absent source below the detection threshold, strong contamination or confusion due to nearby source(s) or bright nebular emission.}

\end{deluxetable}

\clearpage
\clearpage

\begin{deluxetable}{cccccccccc}
\centering \rotate \tabletypesize{\tiny} \tablewidth{0pt}
\tablecolumns{10}
\tablecaption{Membership and Classification of X-ray Sources \label{tbl2}}
\tablehead{

\colhead{No} & \colhead{$\alpha_0$} & \colhead{N$_{\rm b}$} & \colhead{$A_J$}
& \colhead{$M$} & \colhead{Class} & \colhead{F$_1$}& \colhead{F$_2$} & \colhead{Subregion} &
\colhead{F$_3$}\\

& & & (mag) & ($M_{\odot}$) & & & & &\\

(1)&(2)&(3)&(4)&(5)&(6)&(7)&(8)&(9)&(10)}

\startdata
280 & $ -1.49 \pm   0.03$ & 4 &   0.6 &   0.2 & DSK & \nodata & 000 & CepOB3b & 0000\\
281 & $ -2.60 \pm   0.02$ & 4 &   0.5 &   0.5 & NoD & \nodata & 000 & CepOB3b & 0000\\
282 & $ -1.53 \pm   0.02$ & 4 &   0.4 &   0.8 & DSK & \nodata & 000 &    CepB & 0000\\
283 & $ -2.67 \pm   0.03$ & 4 &   0.5 &   0.7 & NoD & \nodata & 000 & CepOB3b & 0000\\
284 & $ -2.65 \pm   0.08$ & 4 &   0.5 &   0.4 & NoD & \nodata & 000 & CepOB3b & 0000\\
285 & $ -1.29 \pm   0.04$ & 4 &   0.6 &   0.3 & DSK & \nodata & 000 &    CepB & 0000\\
286 & $ -1.73 \pm   0.01$ & 4 &   0.5 &   0.5 & DSK & \nodata & 000 &    S155 & 0000\\
287 & $ -2.72 \pm   0.08$ & 3 &   0.1 &   0.5 & NoD & \nodata & 300 &    CepB & 0000\\
288 & $ -2.71 \pm   0.06$ & 3 &   1.0 &   1.0 & NoD & \nodata & 300 &    S155 & 0000\\
289 & $ -2.88 \pm   0.01$ & 4 & \nodata & \nodata & FRG & \nodata & 333 & CepOB3b & 0000\\
290 & $ -2.27 \pm   0.23$ & 2 & \nodata & \nodata & FRG & \nodata & 333 &    CepB & 0000\\
291 & $ -2.41 \pm   0.05$ & 4 &   0.5 &   0.2 & DSK &  TD & 011 & CepOB3b & 0000\\
292 & $ -2.46 \pm   0.06$ & 3 &   0.8 &   0.7 & DSK &  TD & 301 &    S155 & 0000\\
293 & $ -2.78 \pm   0.01$ & 4 &   0.5 &   0.8 & NoD & \nodata & 000 & CepOB3b & 0000\\
294 & $ -2.77 \pm   0.01$ & 4 &   0.7 &   2.1 & NoD & \nodata & 000 & CepOB3b & 0000\\
295 & $ -2.03 \pm   0.17$ & 3 &   0.7 &   0.4 & DSK & \nodata & 300 &    CepB & 0000\\
296 & \nodata & 0 & \nodata & \nodata & UNC & \nodata & 333 &    CepB & 0000\\
\enddata

\tablecomments{This table is available in its entirety in the electronic edition of the journal. A portion is shown here for guidance regarding its form and content. Column 1: X-ray source number. Column 2: SED slope from IRAC photometry with 1$\sigma$ error. Column 3: Number of IRAC bands from which the SED slope was derived. Columns 4-5: $J$-band source extinction and stellar mass estimated from the 2MASS $J$ $vs.$ $J-H$ color-magnitude diagram and a 3(2)~Myr PMS isochrones for the Cep~OB3b/S~155(CepB) regions assuming $d=725$~pc. Column 6: Membership and PMS class. The membership is from \citet{Getman06} and the PMS class is derived here: ``EXG'' -- possible extragalactic contaminant; ``FRG'' -- possible foreground contaminant; ``DSK'' and ``NoD'' -- disk-bearing and diskless PMS stars, respectively; ``UNC'' -- object of the uncertain class.  Column 7: Flag indicating transition disks as based on our visual inspection of SEDs. Column 8: Three-digit flag indicating source position on IR color-color diagrams: $3.6 - 4.5$ vs. $5.8 - 8.0$ diagram, $3.6 - 4.5$ vs. $4.5 - 5.8$ diagram, and $K_s - 3.6$ vs. $3.6 - 4.5$ diagram. Flag values: ``0'' -- the PMS classification from Column 6 is consistent with location in the color-color diagram; ``1'' -- disk-bearing PMS star from Column 6 lies in the locus of diskless stars in the color-color diagram; ``2'' -- diskless PMS candidate from Column 6 lies in the locus of disk-bearing stars in the color-color diagram; ``3'' -- not a PMS member from Column 6 or not enough information to place a member on the corresponding color-color diagram. Column 9: Subregion: ``CepOB3b'' -- lightly absorbed Cepheus OB3b cluster; ``CepB'' -- molecular cloud; ``S155'' -- H~II region interface. Column 10:  Positional flag for each IRAC channel: ``0'' -- source lies within the IRAC channel field of view; ``1'' -- source lies outside the IRAC channel field.}
\end{deluxetable}

\clearpage \clearpage


\begin{deluxetable}{cccccccccccccc}
\centering \rotate \tabletypesize{\tiny} \tablewidth{0pt}
\tablecolumns{14}
\tablecaption{Infrared Properties of non-$Chandra$ IR-Excess Members \label{tbl3}}
\tablehead{

\colhead{No.} & \colhead{R.A.} & \colhead{Decl.} &
\colhead{$J$} & \colhead{$H$} & \colhead{$K_s$} &
\colhead{F$_1$} & \colhead{[3.6]} &
\colhead{[4.5]} & \colhead{[5.8]} &
\colhead{[8.0]} & Subregion & $A_J$ & $M$\\

&(deg)&(deg)&(mag)&(mag)&(mag)&&(mag)&(mag)&(mag)&(mag)&&(mag)&($M_{\odot}$)\\

(1)&(2)&(3)&(4)&(5)&(6)&(7)&(8)&(9)&(10)&(11)&(12)&(13)&(14)}

\startdata
  1 & 343.90431 & 62.619685 & $ 16.26 \pm   0.12$ & $ 14.99 \pm   0.09$ & $ 13.97 \pm   0.06$ & BAAcc0 & $ 12.49 \pm   0.02$ & $ 11.81 \pm   0.02$ & $ 11.15 \pm   0.03$ & $ 10.34 \pm   0.03$ & CepOB3b &   1.3 &   0.20\\
  2 & 343.90796 & 62.613150 & $ 15.67 \pm   0.07$ & $ 14.36 \pm   0.05$ & $ 13.60 \pm   0.04$ & AAA000 & $ 12.41 \pm   0.02$ & $ 11.95 \pm   0.02$ & $ 11.64 \pm   0.03$ & $ 11.06 \pm   0.03$ & CepOB3b &   1.3 &   0.30\\
  3 & 343.90916 & 62.601677 & $ 15.06 \pm   0.04$ & $ 13.76 \pm   0.04$ & $ 12.85 \pm   0.02$ & AAA000 & $ 11.66 \pm   0.02$ & $ 11.12 \pm   0.02$ & $ 10.67 \pm   0.03$ & $  9.92 \pm   0.03$ & CepOB3b &   1.2 &   0.43\\
  4 & 343.91215 & 62.631087 & $ 15.20 \pm   0.05$ & $ 14.21 \pm   0.08$ & $ 13.93 \pm   0.05$ & AAA000 & $ 13.32 \pm   0.03$ & $ 12.98 \pm   0.03$ & $ 12.57 \pm   0.03$ & $ 11.81 \pm   0.03$ & CepOB3b &   0.7 &   0.27\\
  5 & 343.91421 & 62.627495 & $ 14.08 \pm   0.03$ & $ 12.97 \pm   0.03$ & $ 12.29 \pm   0.02$ & AAA000 & $ 11.46 \pm   0.02$ & $ 11.02 \pm   0.02$ & $ 10.74 \pm   0.03$ & $ 10.03 \pm   0.03$ & CepOB3b &   0.7 &   0.61\\
  6 & 343.91472 & 62.634968 & $ 15.22 \pm   0.05$ & $ 14.20 \pm   0.05$ & $ 13.51 \pm   0.04$ & AAA000 & $ 12.66 \pm   0.02$ & $ 12.29 \pm   0.02$ & $ 12.14 \pm   0.03$ & $ 11.69 \pm   0.03$ & CepOB3b &   0.7 &   0.27\\
  7 & 343.92033 & 62.641813 & $ 15.76 \pm   0.07$ & $ 14.58 \pm   0.07$ & $ 13.88 \pm   0.05$ & AAA000 & $ 12.69 \pm   0.02$ & $ 12.28 \pm   0.02$ & $ 11.95 \pm   0.03$ & $ 11.39 \pm   0.03$ & CepOB3b &   1.0 &   0.24\\
  8 & 343.92861 & 62.671552 & $ 15.32 \pm   0.05$ & $ 14.41 \pm   0.06$ & $ 14.06 \pm   0.05$ & AAA000 & $ 13.58 \pm   0.02$ & $ 13.36 \pm   0.02$ & $ 13.22 \pm   0.03$ & $ 12.75 \pm   0.04$ & CepOB3b &   0.5 &   0.22\\
  9 & 343.93061 & 62.645682 & $ 16.03 \pm   0.10$ & $ 14.74 \pm   0.07$ & $ 13.88 \pm   0.06$ & AAAcc0 & $ 12.67 \pm   0.02$ & $ 12.04 \pm   0.02$ & $ 11.38 \pm   0.03$ & $ 10.35 \pm   0.03$ & CepOB3b &   1.3 &   0.24\\
 10 & 343.93414 & 62.631564 & $ 15.21 \pm   0.05$ & $ 14.24 \pm   0.05$ & $ 13.46 \pm   0.03$ & AAA000 & $ 12.18 \pm   0.02$ & $ 11.72 \pm   0.02$ & $ 11.48 \pm   0.03$ & $ 11.00 \pm   0.03$ & CepOB3b &   0.5 &   0.25\\
\enddata

\tablecomments{This table is available in its entirety in the electronic edition of the journal. A portion is shown here for guidance regarding its form and content. Column 1: IR-excess source number. Columns 2-3: IRAC right ascension and declination for epoch J2000.0 in degrees. Columns 4-7: 2MASS $JHK_s$ magnitudes, and 2MASS photometry quality and confusion-contamination flag. Columns 8-11: IRAC magnitudes derived in this work. Column 12: Indicates on which region the source position is projected: ``CepOB3b'' -- lightly absorbed Cepheus OB3b cluster; ``CepB'' -- molecular cloud; ``S155'' -- H II interface. Columns 13-14: $J$-band source extinction and stellar mass derived from the $J$ $vs.$ $J-H$ color-magnitude diagram by dereddening 2MASS photometric colors to the 3(2)~Myr PMS isochrones for the Cep~OB3b/S~155(CepB) regions assuming $d=725$~pc.}

\end{deluxetable}

\clearpage \clearpage



\begin{deluxetable}{ccccccc}
\centering \rotate \tabletypesize{\tiny} \tablewidth{0pt}
\tablecolumns{7}
\tablecaption{Disk Fractions of mass complete PMS samples\label{tbl_frac}}
\tablehead{

\colhead{Class} & \colhead{Cep~OB3b} & \colhead{Cep~OB3b} &
\colhead{S~155} & \colhead{S~155} & \colhead{Cep~B} &
\colhead{Cep~B}\\

&NE&SW&NE&SW&NE&SW\\

(1)&(2)&(3)&(4)&(5)&(6)&(7)}

\startdata
\multicolumn{7}{c}{\bf $0.5 \la M \la 1$~M$_{\odot}$} \\
DSK & 28 & 14 & 10 & 9 & 5 & 4\\
NoD & 34 & 14 & 14 & 4 & 4 & 3\\
Disk Frac. & $0.45_{-0.06}^{+0.06}$ & $0.50_{-0.09}^{+0.09}$ & $0.42_{-0.09}^{+0.10}$ & $0.69_{-0.14}^{+0.10}$ & $0.56_{-0.16}^{+0.14}$ & $0.57_{-0.18}^{+0.15}$\\
\hline
\multicolumn{7}{c}{\bf $M \ga 1$~M$_{\odot}$ } \\
DSK & 10 & 11 & 4 & 3 & 9 & 4\\
NoD & 16 & 13 & 1 & 3 & 2 & 1\\
Disk Frac. & $0.38_{-0.08}^{+0.10}$ & $0.46_{-0.09}^{+0.1}$ & $0.80_{-0.25}^{+0.08}$ & $0.50_{-0.17}^{+0.18}$ & $0.82_{-0.16}^{+0.06}$ & $0.80_{-0.25}^{+0.08}$\\
\hline
\multicolumn{7}{c}{Total over the $M \ga 0.5$~M$_{\odot}$ mass range } \\
DSK & 38 & 25 & 14 & 12 & 14 & 8\\
NoD & 50 & 27 & 15 & 7 & 6 & 4\\
Disk Frac. & $0.43_{-0.05}^{+0.05}$ & $0.48_{-0.07}^{+0.07}$ & $0.48_{-0.09}^{+0.09}$ & $0.63_{-0.12}^{+0.10}$ & $0.70_{-0.11}^{+0.08}$ & $0.67_{-0.15}^{+0.10}$\\
\hline
\hline\\
Class & Inner && Intermediate && Outer &\\
& ( 0\arcmin $-$ 3\arcmin) && (4\arcmin $-$ 7\arcmin) && ($>8$\arcmin) &\\
\hline
\multicolumn{7}{c}{\bf $0.5 \la M \la 1$~M$_{\odot}$} \\
DSK & 9 && 30 && 10 &\\
NoD & 7 && 24 && 20 &\\
Disk Frac. & $0.56_{-0.12}^{+0.11}$ && $0.56_{-0.07}^{+0.06}$ && $0.33_{-0.07}^{+0.10}$ &\\
\hline
\multicolumn{7}{c}{\bf $M \ga 1$~M$_{\odot}$ } \\
DSK & 13 && 17 && 2 &\\
NoD & 3 && 13 && 13 &\\
Disk Frac. & $0.81_{-0.13}^{+0.06}$ && $0.57_{-0.09}^{+0.08}$ && $0.13_{-0.04}^{+0.13}$ &\\
\hline
\multicolumn{7}{c}{Total over the $M \ga 0.5$~M$_{\odot}$ mass range } \\
DSK & 22 && 47 && 12 &\\
NoD & 10 && 37 && 33 &\\
Disk Frac. & $0.69_{-0.09}^{+0.07}$ && $0.56_{-0.05}^{+0.05}$ && $0.27_{-0.05}^{+0.07}$ &\\
\enddata

\end{deluxetable}

\clearpage \clearpage


\clearpage

\clearpage

 
\clearpage

\clearpage

\clearpage

\clearpage

\clearpage


\clearpage

\clearpage


\clearpage

\clearpage

\clearpage

\clearpage

\clearpage

\end{document}